\documentclass{cta-author}
\usepackage{comment}
\usepackage{graphicx}
\usepackage{amssymb}
\usepackage{amsthm}
\usepackage{amsmath}
\usepackage{breqn}
\usepackage{wrapfig}
\usepackage{graphicx}
\usepackage{caption}
\usepackage[english]{babel}
\usepackage{blindtext}
\usepackage{algorithmicx}
\usepackage{algorithm}
\usepackage{algpseudocode}
\usepackage[labelsep=period]{caption}
\usepackage{graphicx}
\usepackage{caption}
\usepackage{subcaption}
\usepackage{multirow}
\usepackage{booktabs}
{}
{}
{}
\usepackage[justification=centering]{caption}
\def\NoNumberAlgo#1{{\def\alglinenumber##1{}\State #1}\addtocounter{ALG@line}{-1}}

\begin{document}

\supertitle{Research Article}

\title{Heuristic Approach for Jointly Optimizing FeICIC and UAV Locations in Multi-Tier LTE-Advanced Public Safety HetNet}

\author{\au{Abhaykumar Kumbhar$^{1,4\corr}$}, \au{Hamidullah Binol $^{2}$}, \au{Simran Singh$^{3}$}, \au{\.{I}smail~G\"uven\c{c}$^{3}$}, \au{Kemal Akkaya$^{1}$}}

\address{\add{1}{Dept. Electrical and Computer Engineering, Florida International University, Miami, FL, USA}
\add{2}{Center for Biomedical Informatics, Wake Forest School of Medicine, Winston-Salem, NC, USA}
\add{3}{Dept. Electrical and Computer Engineering, North Carolina State University, Raleigh, NC, USA}
\add{4}{Motorola Solutions, Inc., Plantation, FL, USA} 
\email{akumb004@fiu.edu}}

\begin{abstract}
Unmanned aerial vehicles (UAVs) enabled communications, and networking can enhance wireless connectivity and support emerging services. However, this would require system-level understanding to modify and extend the existing terrestrial network infrastructure. In this paper, we integrate UAVs both as user equipment and base stations into an existing LTE-Advanced heterogeneous network (HetNet) and provide system-level insights of this three-tier LTE-Advanced air-ground HetNet (AG-HetNet). This AG-HetNet leverages cell range expansion (CRE), intercell interference coordination (ICIC), 3D beamforming, and enhanced support for UAVs. Using this system-level understanding and through brute-force technique and heuristics algorithms, we evaluate the performance of AG-HetNet in terms of fifth percentile spectral efficiency (5pSE) and coverage probability. In particular, we compare system-wide 5pSE and coverage probability, when unmanned aerial base stations (UABS) are deployed on a fixed hexagonal grid and when their locations are optimized using a genetic algorithm (GA) and elitist harmony search algorithm based on the genetic algorithm (eHSGA); while jointly optimizing the ICIC and CRE network parameters for different ICIC techniques. Our simulation results show that the heuristic algorithms (GA and eHSGA) outperform the brute-force technique and achieve better peak values of coverage probability and 5pSE. Simulation results also show that a trade-off exists between peak values and computation time when using heuristic algorithms. Furthermore, the three-tier hierarchical structuring of reduced power subframes FeICIC defined in 3GPP Rel-11 provides considerably better 5pSE and coverage probability than the 3GPP Rel-10 with almost blank subframes eICIC. We also investigate the network performance for different practical deployment heights of UABS, and we find low-altitude UABSs ($25$ m) to perform sparsely better than medium-altitude UABSs ( $36$ m and $50$ m).
\end{abstract}

\maketitle

\section{Introduction}\label{sec1}

Recent developments in reliability and cost-effective hardware have enhanced the drones or unmanned aerial vehicles (UAV) capabilities such as mobility, location-aware connectivity, deployment flexibility in three-dimensional (3D) space, and enabling ubiquitous and non-line-of-sight connectivity.
In particular, UAVs are deployed as unmanned aerial base-stations (UABSs) to meet the mobile data and coverage demands and to restore damaged infrastructure by relieving the pressure on the terrestrial networks and reducing the cost of dense small cell deployments~\cite{R1,chandrasekharan2016designing, merwaday2016improved, kumbhar2018exploiting}. For example, in the aftermath of Hurricane Maria,  AT\&T deployed cell on wings (COW) drone to restore long term evolution (LTE) cell coverage in Puerto Rico~\cite{attCow, drive1, cnbc1}. On the other hand, Verizon has been testing a flying cell site that provides LTE coverage of one-mile range~\cite{cnbc1}. To this end, several of the academic literature has investigated the role of UABSs for improving wireless coverage and spectral efficiency (SE) in~\cite{merwaday2016improved, kumbhar2018exploiting,VZWSmallCell,ATTSmallCell,nakamura2013trends,al2017full, galkin2017stochastic, turgut2018downlink,azari2017coexistence,singh2019distributed}.

The UABS-based communications and networking present research challenges in the field of network planning, optimal 3D deployment, interference management, performance characterization, handover management, and integrating a suitable channel model. However, the existing literature has focused mostly on particular aspects of UABS-based communications and not the air-ground HetNet (AG-HetNet) system as a whole.
In particular,~\cite{naranguav,merwaday2016improved,al2014optimal,bor2016efficient,mozaffari2016optimal,sharma2016uav,christy2017optimum,rupasinghe2019non,sun2018location,hanna2019distributed,challita2019interference,sharma2016uav} have explored UABS location optimization and deployment height, but key aspects such as inter-cell interference coordination (ICIC) techniques and air-ground path loss model are not explicitly taken into account. The effect of interference in a UABS-based network is investigated in~\cite{mozaffari2015drone}, by measuring the optimal distance between the two interfering UABSs and positioning each UABS at a fixed height to maximize the coverage area. Whereas, in~\cite{saquib2013fractional}, a fractional frequency reuse method is used to mitigate interference in a fixed HetNet, to improve the indoor coverage and maximize the network SE by minimizing the user equipment (UE) outage probability. Moreover, a priority-based UE offloading and UE association with mobile small cells for public safety communication (PSC) is studied in~\cite{kaleem2016public}. However,~\cite{mozaffari2015drone,saquib2013fractional,kaleem2016public} did not consider any of the 3GPP Rel-10 and Rel-11 ICIC techniques for the HetNet deployments. 

\begin{figure} [!t]
\centering
\includegraphics[width=1.05\linewidth]{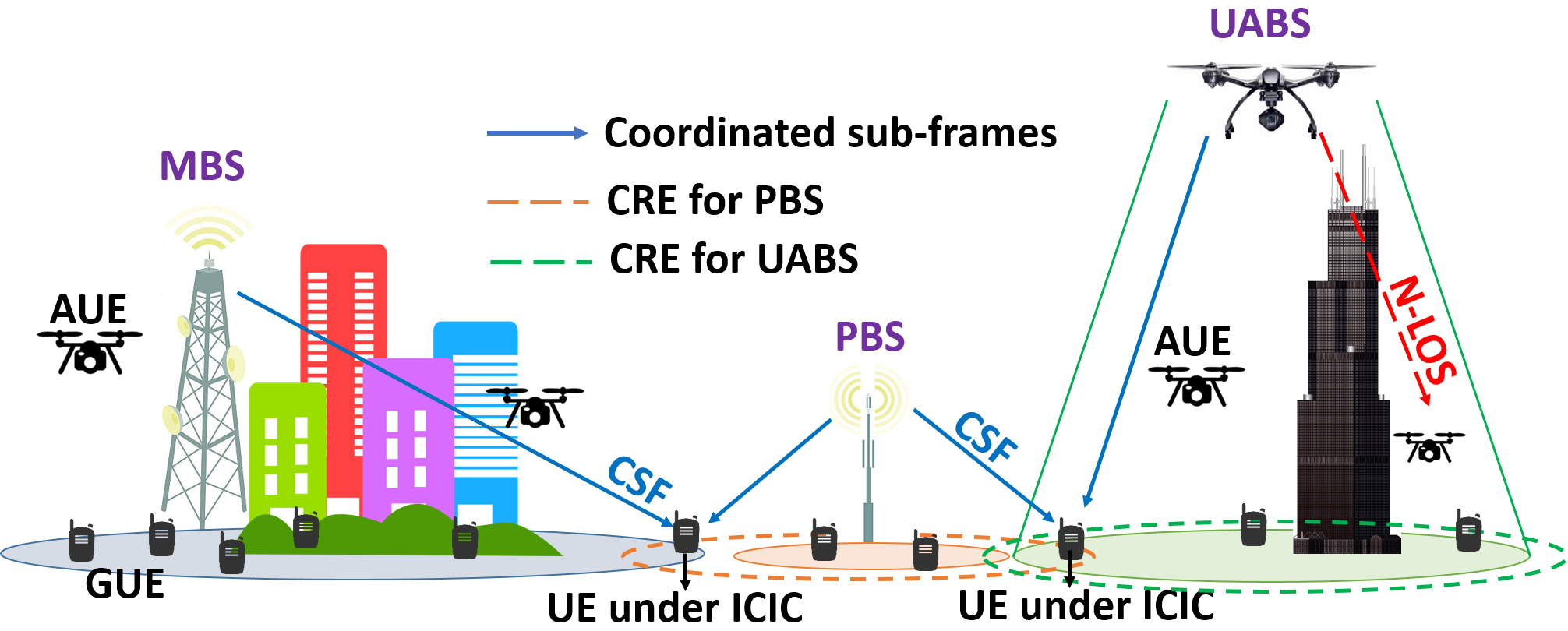}
\caption{The terrestrial nodes (MBS, PBS, and GUE) and aerial nodes (UABS and AUE) constitute the AG-HetNet. The MBS and PBS can use inter-cell interference coordination techniques defined in LTE-Advanced. The PBSs and mobile UABSs can utilize range expansion bias to offload UEs in network congested areas.}
\label{PscHetnet}
\vspace{-2mm}
\end{figure}

The effectiveness of 3GPP Rel-10 enhanced ICIC (eICIC) and Rel-11 further-enhanced ICIC (FeICIC) techniques while taking cell range expansion (CRE) into account has been studied in \cite{kumbhar2018exploiting,singh2019distributed,deb2014algorithms,R10,mukherjee2011effects} for LTE-Advanced HetNet. Authors in~\cite{deb2014algorithms} propose algorithms that jointly optimizes the eICIC parameters, UE cell association rules, and spectrum resources shared between the macro base-stations (MBSs) and fixed small cells. However, in~\citep{deb2014algorithms}, the 3GPP Rel-11 FeICIC technique is not considered, which provides better utilization of radio resources and can offload a larger number of UEs to small cells through CRE. The benefits of 3GPP Rel-10 and Rel-11 ICIC techniques with CRE has been investigated in~\cite{R10}, but for a terrestrial LTE-Advanced HetNet. UABS-assisted LTE-Advanced HetNet has been explored in~\cite{kumbhar2018exploiting,singh2019distributed}, where the UABSs uses CRE for offloading users from a macrocell while considering 3GPP Rel-10 and Rel-11 ICIC techniques in the cell expanded region. Furthermore, a brute-force technique and heuristic algorithm is also used to maximize the spectral efficiency gains by optimizing UABS locations and ICIC network parameters. However, the coverage probability of the wireless network is not investigated, and channel modeling designed for aerial vehicles is not taken into account.

\begin{table*}[!t]
\centering
\captionsetup{justification=centering}
\caption{Literature review on heuristics approach used for joint optimization of interference coordination and UABS locations placement in AG-HetNet.\label{Table:contriSum}}{%
\begin{tabular}{@{}p{2cm} p{2.5cm} p{3.5cm} p{4cm} p{4.75cm}@{}} \toprule
\textbf{Reference} &\textbf{Wireless} & \textbf{Path loss} & \textbf{Optimization} & \textbf{Optimization}\\
 & \textbf{nodes} &\textbf{model} & \textbf{techniques} &  \textbf{goal}\\
\midrule
\citep{merwaday2016improved} & MBS, UABS, GUE & Log distance & Brute-force, Genetic algorithm & UABS locations, spectral efficiency, coverage \\
\citep{sharma2016uav} & MBS, UABS, GUE & Log distance & Neural model & UABS locations \\
\citep{singh2019distributed} & MBS, UABS, GUE & Log distance & Q-learning, Deep Q-learning, Brute-force, Sequential algorithm & UABS locations, spectral efficiency, energy efficiency, interference coordination\\
\citep{kumbhar2018exploiting} & MBS, UABS, GUE & Log distance, Okumura-Hata & Fixed hexagonal, Brute-force, Genetic algorithm & UABS locations, spectral efficiency, energy efficiency, interference coordination \\
\citep{kumbhar2018interference} & MBS, PBS, UABS, GUE, AUE &  Okumura-Hata, ITU-R P.1410-2, 3GPP RP-170779 & Brute-force & UABS locations, spectral efficiency, coverage probability, energy efficiency, interference coordination \\
\citep{rupasinghe2019non} &  UABS, GUE &  Log distance, Close-in mmWave model & - & Spectral efficiency, coverage probability\\
\citep{sun2018location} &  UABS, GUE & ITU-R P.1410-2 & Region partition strategy, Backtracking line search algorithm & UABS locations, GUE load balancing \\
\citep{hanna2019distributed} & UABS swarm & MIMO channel & Brute-force, Gradient descent location optimization & UABS locations, spectral efficiency \\
\citep{challita2019interference} & MBS, GUE, UABS &  ITU-R P.1410-2,3GPP TR 25.942 & Deep reinforcement learning & UABS locations, energy efficiency, wireless latency, interference coordination\\
\citep{zhang2018machine} & UABS, GUE & ITU-R P.1410-2 & Centralized machine learning & UABS locations, energy efficiency \\
\citep{zhang2018predictive} & MBS, UABS, GUE& ITU-R P.1410-2 & Wavelet transform machine learning & UABS locations, GUE load balancing\\
\citep{sharafeddine2019demand} & MBS, GUE, UABS & ITU-R P.1410-2 & Greedy approach  & UABS 3D-locations, GUE load balancing \\
\citep{li2018placement} & UABS, GUE & Free space & Alternating optimization, Successive convex programming & UABS locations, bandwidth allocation, energy Efficiency \\
\citep{fouda2019interference} & UABS, GUE, MBS & MISO channel & Hybrid fixed-point iteration, particle swarm optimization & UABS 3D-locations, coverage probability, interference management, spectral efficiency \\
\textbf{Our work} & \textbf{MBS, PBS, UABS, GUE, AUE} & \textbf{Okumura-Hata, ITU-R P.1410-2, 3GPP RP-170779} & \textbf{Brute-force, Genetic algorithm, eHSGA} & \textbf{UABS locations, spectral efficiency, coverage probability, energy efficiency, interference coordination}\\
\botrule
\end{tabular}}{}
\end{table*}

In a broader context, the use of UAV as aerial user equipment (AUEs) has enabled smart city applications such as traffic monitoring, data collection from Internet-of-Things (IoT) nodes, and public safety applications such as search and rescue, and remote location sensing. In the most recent Kilauea volcano eruption, the first responders were able to search and rescue a Hawaiian man using a UAV~\cite{drive2}. Such vast applications have enabled the recent works to study the feasibility of deploying AUEs in collaboration with existing LTE-Advanced infrastructure in~\cite{menouar2017uav, koohifar2017receding, saputro2018supporting,drive2,niu2018uav,van2016lte,amorim2018measured,kumbhar2018interference}. However, when AUEs are deployed as part of existing terrestrial infrastructure, they experience the same interference issues in the downlink and signal degradation due to path loss. And to address these concerns, a relevant investigation is needed, if AUEs has to coexist with LTE-Advanced HetNet effectively. 

In particular, AUEs applications for smart cities has been studied in~\cite{menouar2017uav,saputro2018supporting,niu2018uav}, and the IoT data are transmitted into LTE base-station or via device-to-device multi-hop communications. However, the coexistence of AUEs with existing terrestrial and aerial nodes is not considered. On the other hand, the coverage probability of AG-HetNet serving AUEs is evaluated while considering appropriate aerial propagation model in~\cite{azari2017coexistence}, but does not consider 3GPP Rel-15 enhanced support for aerial vehicles and the impact of interference on the AUEs in AG-HetNet. The effectiveness of 3GPP Rel-15 enhanced support for aerial vehicles, and the interference mitigation to improve the data capacity is investigated in~\cite{amorim2018measured,van2016lte}, while the AUEs coexist in the AG-HetNet. Nonetheless, the study does not investigate 3GPP Rel-10/11 interference mitigation techniques in the downlink of the AUEs and the influence of CRE on AUEs while mitigating interference.

\subsection{Contributions}
The integration of the UAVs as both AUEs and UABSs would require a system-level understanding to both modify and extend the existing terrestrial network infrastructure. A vital goal while planning any AG-HetNet is to ensure ubiquitous data coverage with broadband rates. To achieve this goal, we define and simulate an AG-HetNet system model for an urban environment with public safety LTE band class~14, as illustrated in Fig.~\ref{PscHetnet}. The proposed AG-HetNet model leverages on 3GPP Rel-8 CRE, 3GPP Rel-10/11 ICIC, 3GPP Rel-12 three-dimensional (3D) beamforming (3DBF), and 3GPP Rel-15 enhanced support for UAVs. Consequently, to assess the performance of this AG-HetNet, we consider \emph{coverage probability} and \emph{fifth percentile SE} (5pSE) as the key performance indicators (KPIs). To maximize the two KPIs of the system model, we jointly optimize the UABS locations in 2D and ICIC and CRE network parameters using a brute-force technique, genetic algorithm (GA), and elitist harmony search algorithm based on the genetic algorithm (eHSGA), while mitigating intercell interference. To reduce the complexity of the optimization algorithms, the deployment height of UABS is not considered during joint optimization. However, we do investigate the impact of UABS height on the overall performance of the wireless network by manually varying the deployment heights. 

To the best of our knowledge, this is the first time in the literature to study the feasibility of deploying UAV as both UABS and AUEs with an existing LTE-Advanced terrestrial infrastructure. Furthermore, the investigation of critical aspects such as the inter-cell interference, channel modeling support, SE, and coverage probability is extended to cover both UABSs and AUEs as part of the AG-HetNet. The specific contributions of our work in the context of the existing literature is summarized in Table~\ref{Table:contriSum}.

The rest of this paper is organized as follows. In Section~\ref{systemModel}, we integrate the 3D channel model and 3DBF with the proposed LTE-Advanced AG-HetNet system model and define the KPIs as a function of network parameters. The UABS deployment and ICIC network parameter optimization using a brute-force technique, GA, and eHSGA are described in Section~\ref{optiDiscuss}. Whereas, in section~\ref{simRes}, through extensive computer simulations, we analyze and compare the two KPIs of the AG-HetNet for various ICIC techniques, deployment heights of UABS, and optimization techniques. Furthermore, we also discuss the impact of UABS height on the KPIs and performance of the optimization techniques. Finally, the last section provides concluding remarks.

\section{System Model}
\label{systemModel}
We consider a three-tier AG-HetNet deployment, where all the MBS, PBS and UABS locations (in 3D) are captured in matrices ${\bf X}_{\rm mbs} \in \mathbb{R}^{N_{\rm mbs}\times 3}$, ${\bf X}_{\rm pbs} \in \mathbb{R}^{N_{\rm pbs}\times 3}$, and ${\bf X}_{\rm uabs}\in \mathbb{R}^{N_{\rm uabs}\times 3}$, respectively, with $N_{\rm mbs}$, $N_{\rm pbs}$ and $N_{\rm uabs}$ denoting the number of MBSs, PBSs, and UABSs within the simulation area (${\rm A_{sim}}$). Similarly, the 3D distribution of GUEs and AUEs are respectively captured in matrices ${\bf X}_{\rm gue}$ and ${\bf X}_{\rm aue}$. Assuming a fixed antenna height, the location of wireless nodes MBS, PBS, GUE, and AUE are modeled using a 2D Poisson point process (PPP), with densities $\lambda_{\rm mbs}$, $\lambda_{\rm pbs}$, $\lambda_{\rm gue}$ and $\lambda_{\rm aue}$, respectively. On the other hand, UABS locations are either optimized using an eHSGA or GA or deployed on a fixed hexagonal grid at low-altitude and medium-altitude~\cite{liu2018performance}. The densities and deployment heights each of the wireless nodes are specified in Table~\ref{tab:SysParams}.

Let $N_{\rm ue}$ be the total numbers of UEs (AUEs + GUEs) to be scheduled, then the nearest distance of an arbitrary $n$th UE from any macrocell of interest (MOI), picocell of interest (MOI), and UABS-cell of interest (UOI) is given by $d_{on}$, $d_{pn}$, and $d_{un}$, respectively. Then assuming Nakagami-m fading channel, the reference symbol received power from MOI, POI, and UOI is given by
\begin{align}
\label{eq:refPwr}  
& {R}_{\rm mbs}(d_{on}) = \frac{P_{\rm mbs}A_E(\phi, \theta)H}{10^{\varphi(d_{on})/10}},   
{R}_{\rm pbs}(d_{pn}) = \frac{P_{\rm pbs}A_E(\phi, \theta)H}{10^{\varphi(d_{pn})/10}}, \nonumber\\ 
& {R}_{\rm uabs}(d_{un}) = \frac{P_{\rm uabs}A_E(\phi, \theta)H}{10^{\varphi(d_{un})/10}},  
\end{align}
where variables $\varphi(d_{on})$, $\varphi(d_{pn})$, and $\varphi(d_{un})$ are path-loss respectively observed from MBS, PBS, and UABS in dB. And random variable $H$ accounts for Nakagami-m fading, whose probability density function is given by~\cite{azari2017coexistence} 
\begin{align}
     f_N(\omega,m)  = \frac{m^m \omega^{m-1}}{\Gamma(m)}\exp(-m\omega),
\end{align}
where $m$ is the shaping parameter, $\omega$ is the channel amplitude and $\Gamma(m)$ is the standard Gamma function given as $\Gamma(m) = \int_{0}^{\infty} \exp(-u)u^{m-1}{\rm d}u$. Through shaping parameter $m$, received signal power can be approximated to variable fading conditions. The value $m > 1$ approximates to Rician fading along line-of-sight (LOS)  and $m = 1$ approximates to Rayleigh fading along non-LOS (NLOS). Furthermore, using the definition of zenith ($\theta$) and azimuth ($\phi$) spherical angles and spherical unit vectors in a Cartesian coordinate, we define variable $A_E(\phi,\theta)$ as the transmitter antenna's 3DBF element and is defined in~\cite{3GPP.TR.36.873dup} as
\begin{align}
     A_E(\phi, \theta) = G_{E,{\max}} - \min\big\{-(A_H(\phi) + A_V(\theta)), A_m \big\}, \\ \ A_m - 30 \ {\rm dB}, \ G_{E,{\max}} = 8 \ {\rm dBi}, \nonumber
     \label{eq:3dbf}
\end{align}
where antenna element for horizontal ($A_H(\phi)$) and vertical ($A_E(\theta)$) radiation pattern, respectively is given by
\begin{align}
     A_H(\phi) = - \min\bigg[ 12\bigg( \frac{\phi}{\phi_{\rm 3dB}}\bigg)^2, A_m \bigg], \ \phi_{\rm 3dB} = 65^\circ,\\
     A_E(\theta) = - \min\bigg[ 12\bigg( \frac{\theta - \theta_{\rm tilt}}{\theta_{\rm 3dB}}\bigg)^2, SLAV \bigg], \ \theta_{\rm tilt} =  90^\circ, \\ \ SLAV = 30, \ \theta_{\rm 3dB} = 65^\circ. \nonumber
\end{align}
Using 3DBF, the power transmission from MBS ($P_{\rm mbs}$), PBS ($P_{\rm pbs}$), and UABS ($P_{\rm uabs}$) can be controlled at UEs in cell-edge/CRE region. Thus limiting the power transmission into adjacent cells which causes inter-cell interference and subsequently improving signal-to-interference ratio (SIR) of the desired signal~\cite{kammoun2014preliminary}.

\subsection{Path Loss Model}
In an urban environment, based on the type of communication link, i.e., ground-to-ground (GTG), any-to-air (ATA), and air-to-ground (ATG) between a UE and base-station (BS) of interest, we consider distinct path-loss models for accurate analysis of signal reliability for the proposed AG-HetNet.

\subsubsection{GTG Communication Link}
We consider Okumura-Hata Path Loss (OHPL) while estimating the GTG communication link between GUE and terrestrial MBS and PBS.  The OHPL is better suited to an urban terrestrial environment, in which the base-station height does not vary ~\cite{kumbhar2018exploiting,xiroOnline} significantly. When a GUE camps on a terrestrial base-station of interest (MOI or POI), OHPL is given by
\begin{dmath}
\varphi(d) =  74.52 + 26.16{\rm log}(f_{\rm c}) - 20.37{\rm log}(h_{\rm bs}) -  3.2({\rm log}(11.75h_{\rm gue}))^2 + 38.35{\rm log}(d), \label{Eq:PlTerrBs} \
\end{dmath}
where $f_{\rm c}$ is the carrier frequency in MHz, $h_{\rm gue}$ is the height of GUE in meter, and $h_{\rm bs}$ is the height of terrestrial base-station in meter i.e., height of MBS is given by $h_{\rm mbs}$ and PBS by $h_{\rm pbs}$.

\subsubsection{ATA Communication Link}
Whenever, an AUE camps on any nearest base-station, we consider a 3D channel model for an urban-macro with aerial (UMa-AV) scenario defined in 3GPP Rel.~15~\cite{3GPP.TR.36.777}. The UMa-AV LOS and NLOS path loss, respectively are given by
\begin{align}
       \varphi(d)  = 
        \begin{cases}
            \varphi^{\rm LOS}(d) = 28.0+22{\rm log} _{10}(d_{\rm 3D}) + 20{\rm log}_{10}(f_{\rm c}) \\
           \varphi^{\rm NLOS}(d) = -17.5+ (46 -7{\rm log}_{10}(h_{\rm aue}))10{\rm log}_{10}(d_{\rm 3D}) \\ \hspace{52pt}+ 20{\rm log}_{10}(\frac{40\pi f_{\rm c}}{3})
        \end{cases}
        , \label{Eq:rel15Pl}
\end{align}
where $f_{\rm c}$ is the carrier frequency in MHz, $d_{\rm 3D}$ is the 3D distance between AUE and the base-station of interest, and $h_{\rm aue}$ is the height of AUE in meter such that $22.5{\rm m} < {\rm h_{aue}} \leq 300{\rm m}$ for $\varphi^{\rm LOS}(d)$ and  
$10.0{\rm m} < {\rm h_{aue}} \leq 100{\rm m}$ for $\varphi^{\rm NLOS}(d)$.

The LOS probabilities for the ATA communication link defined in 3GPP Rel.~15~\cite{3GPP.TR.36.777} is given 
\begin{equation}
      \mathbb{P}_{\rm LOS}(\varphi) = 
        \begin{cases}
            1,  & d_{\rm 2D} \leq d_1 \\
            \frac{d_1}{d_{\rm 2D}} + \exp(\frac{-d_{\rm 2D}}{p_1})(1-\frac{d_1}{d_{\rm 2D}}), & d_{\rm 2D} > d_1
        \end{cases}
        ,\label{Eq:rel15PlProb}
 \end{equation}
where is $d_{\rm 2D}$ is the 2D distance between AUE and the base-station of interest such that $d_{\rm 2D} \leq 4{\rm km}$, and the factors $p_1$ and $d_1$ (in meters) are given by 
\begin{align}
&  p_1 = 4300{\rm log}_{10}(h_{\rm aue})-3800, \nonumber \\ 
&  d_1 = \max(460{\rm log}_{10}(h_{\rm aue})-700, 18). \nonumber
\end{align}

Using this model, we calculate the average path loss over the probabilities of LOS and NLOS communication link between AUE and the camping base-station. Then using \eqref{Eq:rel15Pl} and \eqref{Eq:rel15PlProb}, the average path loss is given by  
\begin{align}
    {\rm PL}_{\rm avg} = \mathbb{P}_{\rm LOS}\times\varphi^{\rm LOS} + (1-\mathbb{P}_{\rm LOS})\times\varphi^{\rm NLOS}.
\end{align}

\begin{figure} [t]
\centering
\includegraphics[width=0.7\linewidth]{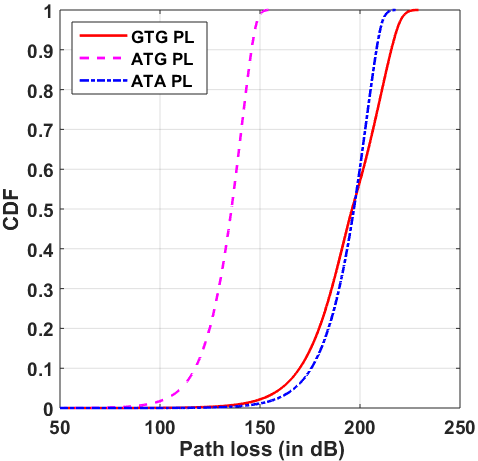}
\caption{The CDF of path loss observed for the communication link between UEs and base-stations.}
\label{fig:all3PL}
\vspace{-2mm}
\end{figure}

\subsubsection{ATG Communication Link}
Whenever, a GUE camps on a UOI, we consider a more conventionally used LOS/NLOS path loss model defined in the literature~\cite{ITU-R.P.1410-2,al2014optimal,khawaja2018survey} and is given by
\begin{align}
&   \varphi_{\rm uabs}(d)  =  \prod_{x = 0}^{y}\Bigg[ 1- \exp \bigg(-\frac{[h_{\rm uabs} - \frac{(x +1/2)(h_{\rm uabs} - h_{\rm gue})}{y+1}]^2}{2\Omega^2}\bigg) \Bigg], \label{Eq:uabsPl}
\end{align}
where $h_{\rm uabs}$ is the deployment height of UABS, $y \ = \ {\rm floor}(r \sqrt[]{\zeta \xi} - 1)$, r is the ground distance between the UABS and GUE, $\zeta$ is the ratio of built-up land area to the total land area, $\xi$ is the mean number of buildings per unit area (buildings/${\rm km}^2$), and $\Omega$ characterizes the building height (denoted by $H_{\rm B}$) distribution  in meters and is based on a Rayleigh distribution: $f(H_{\rm B}) = \frac{H_{\rm B}}{\Omega^2}\exp(\frac{-H_{\rm B}^2}{2\Omega^2})$. Furthermore, we consider LOS probability $\mathbb{P}_{\rm LOS}(\varphi_{\rm uabs})$ as a continuous function of $\theta$ and environment factors. By approximating environment factors to a simple modified Sigmoid function (S-curve), the simplified LOS probability is given by
\begin{align}
&   \mathbb{P}_{\rm LOS}(\varphi_{\rm uabs}, \theta)  =   \frac{1}{1 + a \exp(-b[\theta - a])},  \label{Eq:uabsPlProb}
\end{align}
where $a$ and $b$ are the S-curve parameters.

Fig.~\ref{fig:all3PL} illustrates the empirical path loss cumulative distribution functions (CDFs), calculated for all distances between base stations (${\bf X}_{\rm mbs}$, ${\bf X}_{\rm pbs}$, and ${\bf X}_{\rm uabs}$) and UEs (${\bf X}_{\rm gue}$ and ${\bf X}_{\rm aue}$), using conditions defined in previous paragraph. Inspection of Fig.~\ref{fig:all3PL} reveals that the maximum allowable path loss is diverse for GTG, ATG, and ATA communication links. This variation is primarily due to the environmental factors and LOS/NLOS probability of communication link. Nevertheless, maximum allowable path-loss for the models used in GTG, ATA, and ATG link is approximately 255 dB, 216 dB, and 154 dB, respectively.

\begin{figure}[!t]
\centering
\includegraphics[width=1.1\linewidth]{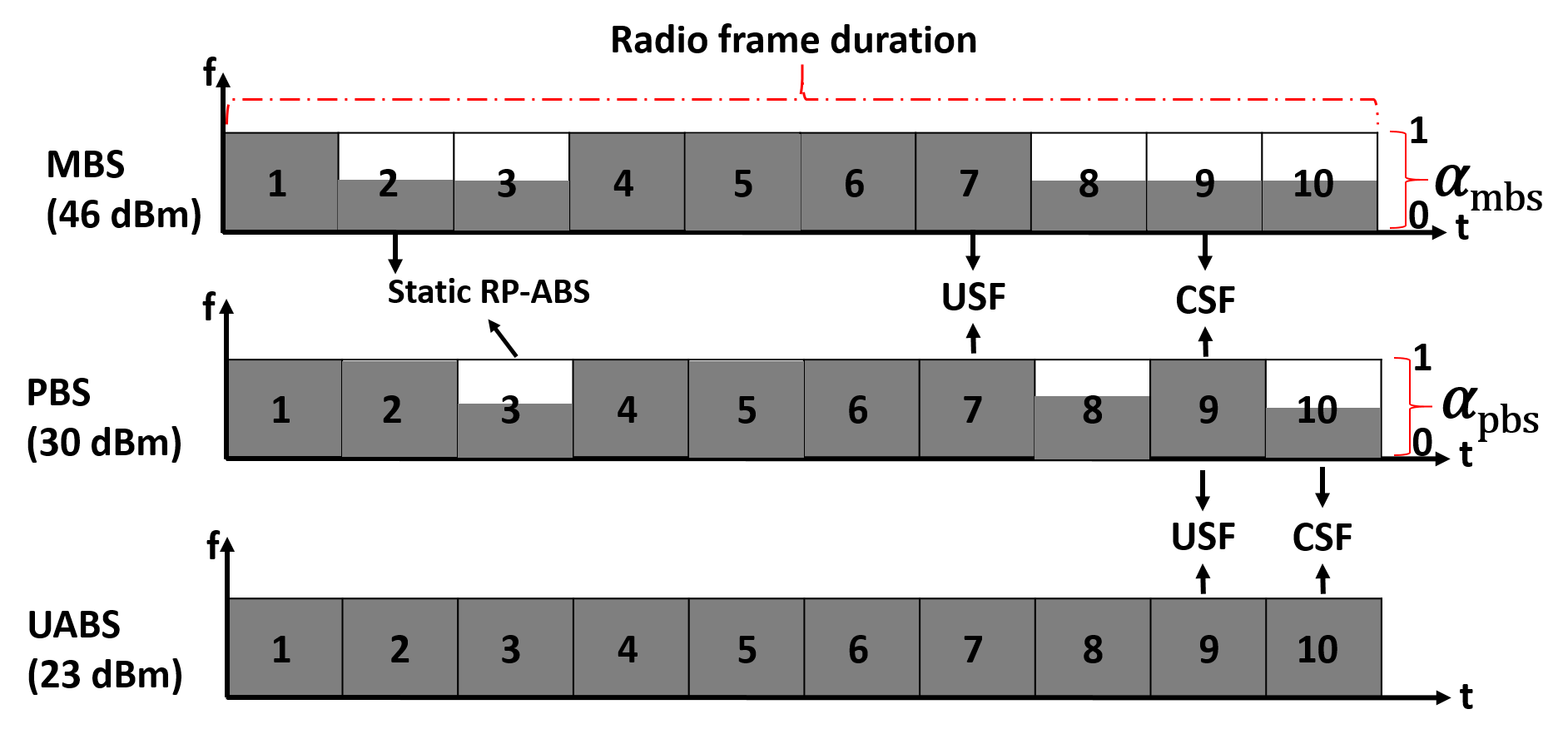}
\caption{The proposed three-tier reduced power USF/CSF LTE subframes of MBS, PBS, and UABS. Certain UABS subframes are protected from both MBS and PBS, while certain PBS subframes are protected from MBS.}
\label{fig:icicRadioFrame}
\vspace{-2mm}
\end{figure}

\begin{figure}[!t]
\centering{\includegraphics[width=1.05\linewidth]{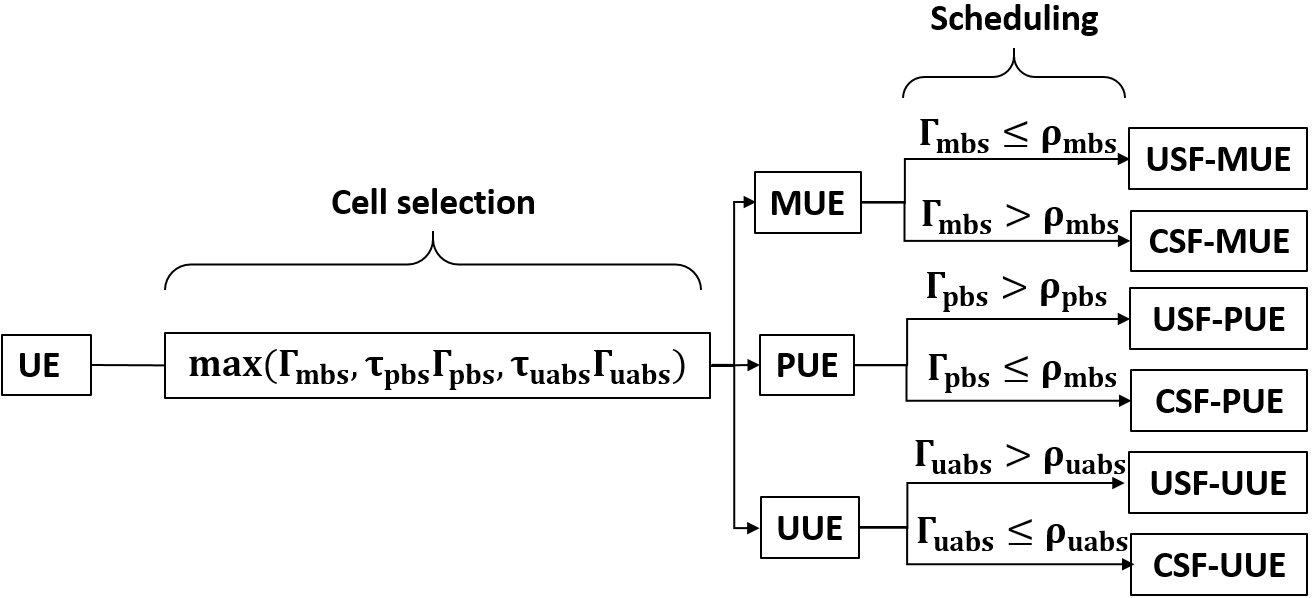}}
\caption{Cell selection and UE association in USF/CSF subframes of MBS, PBS, and UABS.}
\label{fig:CellSelection}
\end{figure}

\subsection{Spectral Efficiency with 3GPP Rel.10/11 ICIC}
\label{icicidetails}
Due to their low transmission power, the small cells such as the PBS and UABS are unable to associate a substantial number of UEs compared to that of MBSs. Therefore, we consider the CRE technique defined in 3GPP Rel-8 at small cells to extend the network coverage and increase capacity by offloading traffic from congested cells. Although an adverse side effect of CRE includes increased interference at UEs in the cell-edge or CRE region. To address this intercell interference, both MBS and PBS are capable of using ICIC techniques defined in 3GPP Rel-10/11~\cite{kumbhar2018exploiting}, wherein both MBS and PBS can transmit radio frames at reduced power levels as shown in Fig.\ref{fig:icicRadioFrame}.

The radio subframes with reduced power are termed as coordinated subframes (CSF) and full power as uncoordinated subframes (USF). The power reduction factor of radio subframes at MBS is given by $\alpha_{\rm mbs}$ and $\alpha_{\rm pbs}$ at PBS. In particular, $\alpha_{\rm mbs}=\alpha_{\rm pbs}=0$ corresponds to Rel-10 almost blank subframes (ABS) eICIC, $\alpha_{\rm mbs}=\alpha_{\rm pbs}=1$ corresponds to no ICIC, and otherwise corresponds to reduced power FeICIC defined in Rel-11. As illustrated in Fig.\ref{fig:icicRadioFrame}, using reduced power FeICIC, we protect certain UABS subframes from both MBS and PBS, while certain PBS subframes are protected from MBS. We coordinate the USF/CSF duty cycle using $\beta_{\rm mbs}$ and $(1-\beta_{\rm mbs})$ at MBS and $\beta_{\rm pbs}$ and $(1-\beta_{\rm pbs})$ at PBS. The proposed AG-HetNet model assumes that the power reduction pattern and radio subframes duty cycle is shared via the X2 interface, which is a logical interface between the base-stations.

Although applying the ICIC technique at each base-station reduces the intercell interference with adjacent cells, it also reduces the desired SIR at the scheduled UEs. Therefore, to improve the desired SIR, we consider the 3DBF at each transmitting base-station to restrict the beamforming and power transmission to the location of scheduled UE~\cite{kammoun2014preliminary}. 

\begin{table*}[!t]
\begin{center}
\captionsetup{justification=centering}
\caption{Signal-to-interference ratio and spectral efficiency definitions.\label{tab:sirCap}}{%
\begin{tabular}{@{}p{6.5cm}|p{6cm}@{}} \toprule
      \textbf{Signal-to-interference ratio} & \textbf{SE in USF/CSF radio frames}\\
      \hline
      $\Gamma^{\rm mbs}_{\rm usf} = \frac{R_{\rm mbs}(d_{on})}{R_{\rm pbs}(d_{pn}) + R_{\rm uabs}(d_{un}) + \mathbf{I}_{\rm agg}}$ & $C_{\rm usf}^{\rm mbs} = \frac{\beta_{\rm mbs} {\rm log_2}(1+\Gamma^{\rm mbs}_{\rm usf})}{N_{\rm usf}^{\rm mbs}}$\\
      $\Gamma^{\rm mbs}_{\rm csf} = \frac{\alpha R_{\rm mbs}(d_{on})}{\alpha_{\rm pbs}R_{\rm pbs}(d_{pn}) + R_{\rm uabs}(d_{un}) +  \mathbf{I}_{\rm agg}}$ & $C_{\rm csf}^{\rm mbs} = \frac{(1-\beta_{\rm mbs}){\rm log_2}(1+\Gamma^{\rm mbs}_{\rm csf})}{N_{\rm csf}^{\rm mbs}}$\\ \hline
      $\Gamma^{\rm pbs}_{\rm usf} = \frac{R_{\rm pbs}(d_{pn})}{R_{\rm mbs}(d_{on}) + R_{\rm uabs}(d_{un}) +  \mathbf{I}_{\rm agg}}$ & $C_{\rm usf}^{\rm pbs} = \frac{\beta_{\rm pbs} {\rm log_2}(1+\Gamma^{\rm pbs}_{\rm usf})}{N_{\rm usf}^{\rm pbs}}$\\
      $\Gamma^{\rm pbs}_{\rm csf} = \frac{\alpha_{\rm pbs} R_{\rm pbs}(d_{pn})}{\alpha R_{\rm mbs}(d_{on}) + R_{\rm uabs}(d_{un})+ \mathbf{I}_{\rm agg}}$ & $C_{\rm csf}^{\rm uabs} = \frac{(1 - \beta_{\rm pbs}) {\rm log_2}(1+\Gamma^{\rm uabs}_{\rm csf})}{N^{\rm pue}_{\rm csf}}$\\ \hline
      $\Gamma^{\rm uabs}_{\rm usf} = \frac{R_{\rm uabs}(d_{un})}{R_{\rm mbs}(d_{on}) + R_{\rm pbs}(d_{pn}) +  \mathbf{I}_{\rm agg}}$ & $C_{\rm usf}^{\rm mbs} = \frac{(\beta_{\rm mbs}+\beta_{\rm pbs}) {\rm log_2}(1+\Gamma^{\rm uabs}_{\rm usf})}{N_{\rm usf}^{\rm uue}}$\\
      $\Gamma^{\rm uabs}_{\rm csf} = \frac{R_{\rm uabs}(d_{un})}{\alpha R_{\rm mbs}(d_{on})+ \alpha_{\rm pbs}R_{\rm pbs}(d_{pn})+  \mathbf{I}_{\rm agg}}$ & $C_{\rm csf}^{\rm uabs} = \frac{(2-(\beta_{\rm mbs} + \beta_{\rm pbs})){\rm log_2}(1+\Gamma^{\rm uabs}_{\rm csf})}{N^{\rm uue}_{\rm csf}}$ \tabularnewline\bottomrule[1pt]%
\end{tabular}}{} 
\end{center}
\vspace{-2mm}
\end{table*}

Given the ICIC framework in 3GPP LTE-Advanced and using a three-tier reduced power USF/CSF structure given in Fig.~\ref{fig:icicRadioFrame}, we define the SIR experienced by a $n$th arbitrary UE scheduled in USF/CSF of MOI, POI, and UOI by following an approach similar to that given in~\cite{kumbhar2018exploiting}. Then, let $\Gamma^{\rm mbs}_{\rm usf}$, $\Gamma^{\rm mbs}_{\rm csf}$, $\Gamma^{\rm pbs}_{\rm usf}$, $\Gamma^{\rm pbs}_{\rm csf}$, $\Gamma^{\rm uabs}_{\rm usf}$, and $\Gamma^{\rm uabs}_{\rm csf}$ be the SIRs for the UE scheduled in the USF/CSF of MOI, POI, and UOI, respectively and is defined in Table~\ref{tab:sirCap}. Wherein, $\mathbf{I}_{\rm agg}$ is the aggregate interference at the UE from all the base-stations, except the MOI, POI, and UOI.

The cell selection process relies on the definition of MOI, POI, and UOI SIRs given in Table~\ref{tab:sirCap}, as well as the CRE $\tau_{\rm pbs}$ at PBSs and $\tau_{\rm uabs}$ at UABSs. Using positive biased CRE $\tau_{\rm pbs}$ at PBSs and $\tau_{\rm uabs}$ at UABSs, the small cells can further expand their SIR coverage. Consequently, during the cell selection process, a UE always camps on an MOI/POI/UOI that yields the best SIR. After cell selection, an MBS-UE (MUE), PBS-UE (PUE), and UABS-UE (UUE) would be scheduled in either USF/CSF radio subframes based on the scheduling threshold of MBS ($\rho_{\rm mbs}$)), PBS ($\rho_{\rm pbs}$), and UABS ($\rho_{\rm uabs}$). This strategy of cell selection and UE scheduling in USF/CSF of MOI/POI/UOI is similar to that of~\cite{kumbhar2018exploiting} and is summarized in Fig.~\ref{fig:CellSelection}. 

Once the $n$th arbitrary UE is assigned to an MOI/POI/UOI and scheduled within the USF/CSF, then using the SIR definitions, the SE of a UE scheduled in the three-tier USF/CSF subframes is defined by $C_{\rm usf}^{\rm mbs}$, $C_{\rm csf}^{\rm mbs}$, $C_{\rm usf}^{\rm pbs}$, $C_{\rm csf}^{\rm pbs}$, $C_{\rm usf}^{\rm uabs}$, and $C_{\rm csf}^{\rm uabs}$ and is given in Table~\ref{tab:sirCap}. Where $N_{\rm usf}^{\rm mue}$, $N_{\rm csf}^{\rm mue}$, $N_{\rm usf}^{\rm pue}$, $N_{\rm csf}^{\rm pue}$, $N^{\rm uue}_{\rm usf}$, and $N^{\rm uue}_{\rm csf}$ are the number of MBS-UE, PBS-UE, and UABS-UE scheduled in USF/CSF of the MBS/PBS/UABS.

\section{UABS Locations and ICIC Parameter Optimization in AG-HetNet}
\label{optiDiscuss}
In this article, 5pSE corresponds to the worst fifth percentile UE capacity amongst all of the scheduled UEs. On the other hand, we define the coverage probability of the network as the percentage of an area having broadband rates and capacity larger than a threshold of~$T_{C_{\rm SE}}$. The primary goal of this simulation study is to maximize these two KPIs while obtaining the best fit ICIC network configuration and optimal UABS locations using a \textit{brute force algorithm}, \textit{genetic algorithm}, and \textit{elitist harmony search based on the genetic algorithm}. 

\begin{figure*}[t]
\centering
\begin{subfigure}[b]{0.47\textwidth}
\label{fig:3dHexDeploy}
\includegraphics[width=0.9\textwidth]{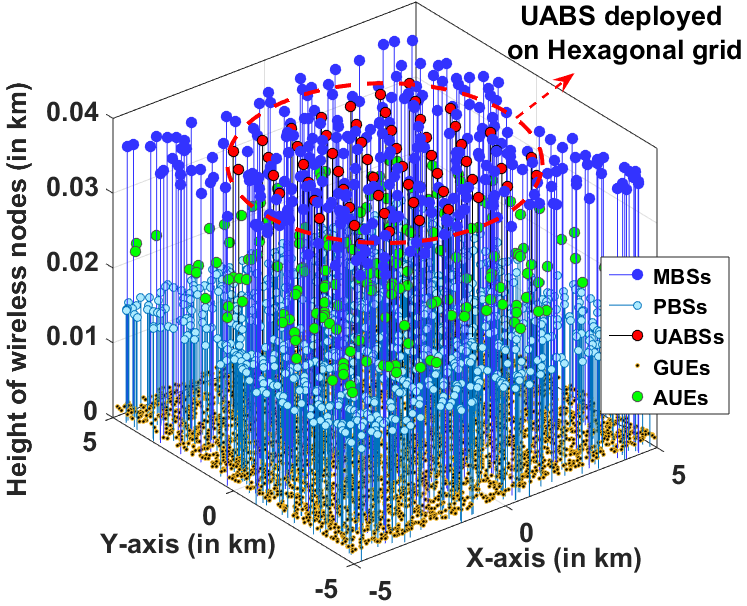}
\caption{UABSs deployed on a fixed hexagonal grid.}
\end{subfigure}
\begin{subfigure}[b]{0.50\textwidth}
\label{fig:3dHeuristicDeploy}
\includegraphics[width=0.9\textwidth]{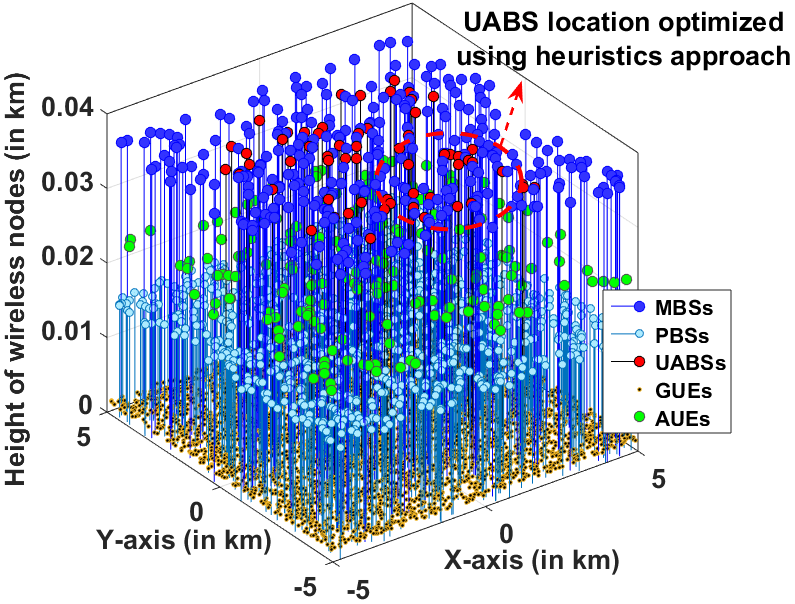}
\caption{UABS locations optimized using heuristics approach.}
\end{subfigure}
\caption{Three dimensional distribution of ground UEs (GUEs), aerial UEs (AUEs), macro base-stations (MBSs), pico base-stations (PBSs), and unmanned aerial base-stations (UABSs) in AG-HetNet. The densities and deployment heights each of the wireless nodes are specified in Table~\ref{tab:SysParams}.}
\label{fig:wirelessNodes} 
\end{figure*}

Consider individual locations $(x_i,y_i)$ of each UABS $i\in\{1,2,...,N_{\rm uabs}\}$ and ${\bf X}_{\rm uabs}$ would be the matrix representing these locations in 3D deployed over a geographical area of interest. The UABSs are placed within the rectangular simulation area regardless of the existing MBS (${\bf X}_{\rm mbs}$) and PBS locations (${\bf X}_{\rm pbs}$). Given the locations of base-station (${\bf X}_{\rm mbs}$, ${\bf X}_{\rm pbs}$, and ${\bf X}_{\rm uabs}$), we capture individual ICIC parameters for each MBS in a matrix ${\bf S}_{\rm mbs}^{\rm ICIC} = [\boldsymbol{\alpha_{\rm mbs}}, \boldsymbol{\beta_{\rm mbs}}, \boldsymbol{\rho_{\rm mbs}}]$ $\in \mathbb{R}^{N_{\rm mbs} \times 3}$, individual ICIC parameters for each PBS in matrix ${\bf S}_{\rm pbs}^{\rm ICIC} = [\boldsymbol{\alpha_{\rm pbs}}, \boldsymbol{\beta_{\rm pbs}}, \boldsymbol{\rho_{\rm pbs}}, \boldsymbol{\tau_{\rm pbs}}]$ $\in \mathbb{R}^{N_{\rm pbs} \times 4}$, and ${\bf S}_{\rm uabs}^{\rm ICIC} = [\boldsymbol{\tau_{\rm uabs}},\boldsymbol{\rho_{\rm uabs}}]$ $\in \mathbb{R}^{N_{\rm uabs} \times 2}$ is a matrix that captures individual ICIC parameters for each UABS. The vectors $\boldsymbol{\alpha_{\rm mbs}}=[\alpha_1,...,\alpha_{N_{\rm mbs}}]^T$, $\boldsymbol{\beta_{\rm mbs}}=[\beta_1,...,\beta_{N_{\rm mbs}}]^T$, and $\boldsymbol{\rho_{\rm mbs}} = [\rho_1,...,\rho_{N_{\rm mbs}}]^T $ capture the power reduction factors, USF duty cycle, and scheduling thresholds, respectively, for each MBS. On the other hand, for each PBS, $\boldsymbol{\alpha_{\rm pbs}}=[\alpha_1,...,\alpha_{N_{\rm pbs}}]^T$, $\boldsymbol{\beta_{\rm pbs}}=[\beta_1,...,\beta_{N_{\rm pbs}}]^T$, $\boldsymbol{\rho_{\rm pbs}} = [\rho_1,...,\rho_{N_{\rm pbs}}]^T $, and $\boldsymbol{\tau_{\rm pbs}} = [\tau_1,...,\tau_{N_{\rm pbs}}]^T $ capture the power reduction, USF duty cycle, scheduling threshold, and range expansion, respectively. Whereas, $\boldsymbol{\rho_{\rm uabs}} = [\rho_1,...,\rho_{N_{\rm uabs}}]^T$ and $\boldsymbol{\tau_{\rm uabs}} = [\tau_1,...,\tau_{N_{\rm uabs}}]^T$ capture the scheduling threshold and range expansion, respectively, for each UABS. Using these variable definitions, the initial state of the AG-HetNet can be given as
$\mathbf{S} = \Big[{\bf X}_{\rm uabs},{\bf S}_{\rm mbs}^{\rm ICIC},{\bf S}_{\rm pbs}^{\rm ICIC},{\bf S}_{\rm uabs}^{\rm ICIC}\Big]$. However, to reduce the system complexity and simulation runtime, we apply the same ${\bf S}_{\rm mbs}^{\rm ICIC}$ parameters across all MBSs, ${\bf S}_{\rm pbs}^{\rm ICIC}$ across all PBSs, and ${\bf S}_{\rm uabs}^{\rm ICIC}$ across all UABSs.

\begin{algorithm}[!t]
    \caption{Brute force algorithm}\label{alg:bruteForceAlgo}
    \begin{algorithmic}[1]
        \Procedure{$C_{\rm \mathbf{KPI}}$}{${\bf X}_{\rm uabs},{\bf S}_{\rm mbs}^{\rm ICIC},{\bf S}_{\rm pbs}^{\rm ICIC},{\bf S}_{\rm uabs}^{\rm ICIC}$}
        \State \texttt{${\rm \mathbf{KPI}}$, Best state $\bf{S^\prime} \leftarrow NULL$}
        \ForAll{\texttt{Values of State $\bf{S}$}}
            \State \texttt{Current KPI $\leftarrow C_{\rm \mathbf{KPI}}(\bf{S})$}
            \If{\texttt{Current KPI $>$ ${\rm \mathbf{KPI}}$}} 
                \State \texttt{${\rm \mathbf{KPI}}$ $\leftarrow$ Current KPI}
                \State \texttt{$\bf{S^\prime} \leftarrow {\bf{S}}$}
            \EndIf
        \EndFor
        \State \texttt{Return ${\rm \mathbf{KPI}}$, Best state $\bf{S^\prime}$}
        \EndProcedure
    \end{algorithmic}
    \label{Alg:bruteForceAlgo}
\end{algorithm}

Using a brute-force algorithm to search for all possible optimal values in a large search space is computationally infeasible. Therefore the UABSs are initially deployed on a fixed hexagonal grid, as shown in Fig.~\ref{fig:wirelessNodes}(a), and every UABS sends its locations and all the base-stations send the spectral efficiency information of its users (AUE and GUE) to a centralized server. Subsequently, a brute-force technique will be used to optimize the only the ICIC network parameters and evaluate the 5pSE and coverage probability for this fixed AG-HetNet. Then, a centralized server can run any appropriate heuristic algorithm to jointly optimize the UABS locations and ICIC parameters, as illustrated in Fig.~\ref{fig:wirelessNodes}(b). Then, for the proposed AG-HetNet the best state ($\bf{S^\prime}_{\mathbf{KPI}}$) of all the possible states, $\bf{S}$ is given as 
\begin{equation}
\mathbf{S^\prime}_{\rm \mathbf{KPI}}  ={\rm \arg}~\underset{\mathbf{S}}{\rm max} ~~ C_{\rm \mathbf{KPI}}(\mathbf{S}),  \label{Eq:optimizeState}
\end{equation}
where $C_{\rm \mathbf{KPI}}(.)$ is an objective function wherein $\mathbf{KPI} \in \big(\rm 5pSE, COV\big)$ then $C_{\rm 5pSE}(.)$ denotes the objective function for 5pSE and $C_{\rm cov}(.)$ denotes the objective function for coverage probability. 

Using the brute-force technique described in Algorithm~\ref{Alg:bruteForceAlgo} and UABS on a fixed hexagonal grid (${\bf X}_{\rm uabs}$) in a AG-HetNet, the optimal values of the best state ($\bf{S^\prime}_{\mathbf{KPI}}$) of all the possible states $\bf{S}$ can be vectorized into $\mathbf{S^\prime}_{\rm \mathbf{KPI}} = \Big[{\bf X}_{\rm uabs},{\bf S^\prime}_{\rm mbs}^{\rm ICIC},{\bf S^\prime}_{\rm pbs}^{\rm ICIC},{\bf S^\prime}_{\rm uabs}^{\rm ICIC}\Big]$. Whereas, using heuristics algorithm proposed in Algorithm~\ref{Alg:GaAlgo} and Algorithm~\ref{Alg:eHsgaAlgo}, the the optimal values of the best state ($\bf{S^\prime}_{\mathbf{KPI}}$) of all the possible states $\bf{S}$ can be vectorized into $\mathbf{S^\prime}_{\rm \mathbf{KPI}} = \Big[{\bf X^\prime}_{\rm uabs},{\bf S^\prime}_{\rm mbs}^{\rm ICIC},{\bf S^\prime}_{\rm pbs}^{\rm ICIC},{\bf S^\prime}_{\rm uabs}^{\rm ICIC}\Big]$. Where, ${\bf X^\prime}_{\rm uabs}$ is the optimal UABS location and ${\bf S^\prime}_{\rm mbs}^{\rm ICIC},{\bf S^\prime}_{\rm pbs}^{\rm ICIC}$, and ${\bf S^\prime}_{\rm uabs}^{\rm ICIC}$ are the optimal ICIC values for MBSs, PBSs, and UABSs, respectively.

\begin{figure} [t]
\centering
\includegraphics[width=1.025\linewidth]{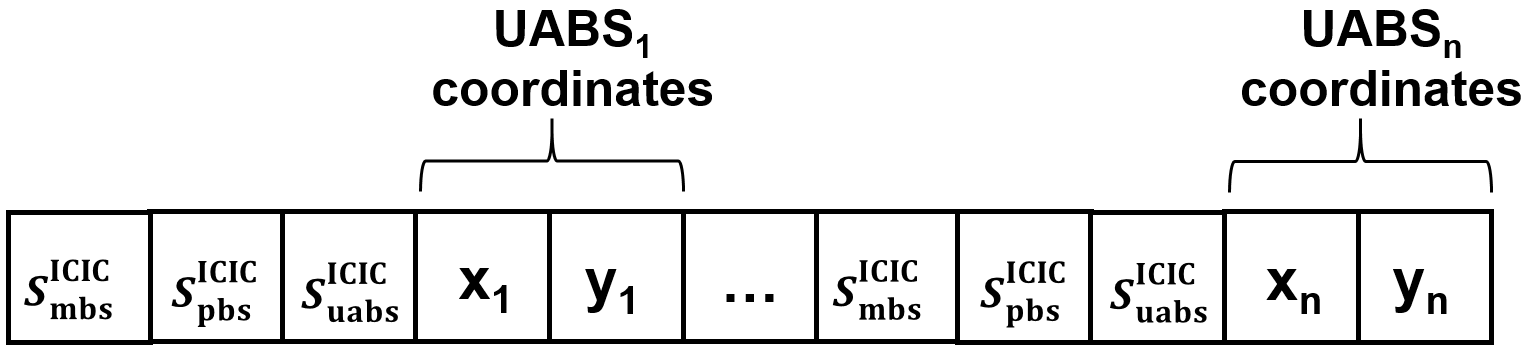}
\caption{An example of a chromosome for ICIC simulation, where the UABS locations and  ICIC parameters ${\bf S}_{\rm mbs}^{\rm ICIC},{\bf S}_{\rm pbs}^{\rm ICIC},$ and ${\bf S}_{\rm uabs}^{\rm ICIC}$ are optimized.}
\label{GAChromosome}
\vspace{-2mm}
\end{figure}

\begin{algorithm}[t]
    \caption{Genetic Algorithm}\label{Alg:GaAlgo}
    \begin{algorithmic}[1]
    \Procedure{$C_{\rm \mathbf{KPI}}$}{${\bf X}_{\rm uabs},{\bf S}_{\rm mbs}^{\rm ICIC},{\bf S}_{\rm pbs}^{\rm ICIC},{\bf S}_{\rm uabs}^{\rm ICIC}$}
        \State \texttt{${\rm \mathbf{KPI}}$, Best state $\bf{S^\prime} \leftarrow NULL$}
        \State \texttt{Selection $\leftarrow$ Roulette Wheel}
        \State \texttt{Initialize genetic parameters:} 
        \NoNumberAlgo{$SZ_{\rm GA}$, $mr$, and $cxr$} 
        \State \texttt{Population (\textbf{POP}) Set of
        \NoNumberAlgo{$\bf{S}$ $\leftarrow$ ${\bf X}_{\rm uabs},{\bf S}_{\rm mbs}^{\rm ICIC},{\bf S}_{\rm pbs}^{\rm ICIC},{\bf S}_{\rm uabs}^{\rm ICIC}$}}
        \State \texttt{FITNESS = $C_{\rm \mathbf{KPI}}(.)$}
        \State \texttt{Evaluate \textbf{POP} FITNESS} 
        \State \texttt{Stop Condition $\leftarrow $ number of iterations}
        \While {\texttt{!Stop Condition}}
            \For{\texttt{$k=1:SZ_{\rm GA}$}}
                \State \texttt{Parent1 $\leftarrow$ SELECTION(\textbf{POP}, FITNESS)}
                \State \texttt{Parent2 $\leftarrow$ SELECTION(\textbf{POP},FITNESS)}
                \State \texttt{Child1, Child2 $\leftarrow$}
                \NoNumberAlgo{\texttt{REPRODUCE(Parent1,Parent2, $cxr$)}}
                \If{$rand() < mr$}
                    \State \texttt{Children <- MUTATE(Child1, $mr$)}
                    \State \texttt{Children <- MUTATE(Child2, $mr$)}
                \EndIf
            \EndFor
            \State \texttt{Evaluate Children FITNESS}
            \State \texttt{Pick best state $\bf{S^\prime}$ from Children}
            \State \texttt{\textbf{POP} $\leftarrow$ REPLACE(\textbf{POP}, Children)}
        \EndWhile
        \State \texttt{Return ${\rm \mathbf{KPI}}$, Best state $\bf{S^\prime}  \leftarrow$ Maximum FITNESS}
    \EndProcedure
    \end{algorithmic}
    \vspace{-1mm}
\end{algorithm}

\begin{algorithm}[t]
    \caption{Elitist Harmony Search-Genetic Algorithm (eHSGA)}\label{Alg:eHsgaAlgo}
    \begin{algorithmic}[1]
        \Procedure{$C_{\rm \mathbf{KPI}}$}{${\bf X}_{\rm uabs},{\bf S}_{\rm mbs}^{\rm ICIC},{\bf S}_{\rm pbs}^{\rm ICIC},{\bf S}_{\rm uabs}^{\rm ICIC}$}
            \State \texttt{${\rm \mathbf{KPI}}$, Best state $\bf{S^\prime} \leftarrow NULL$}
            \State \texttt{Selection $\leftarrow$ Roulette Wheel}
            \State \texttt{Initialize harmony search parameters:} 
            \NoNumberAlgo{$SZ_{\rm HM}$, $par$, $R_{HMC}$, and $fr$} 
            \State \texttt{Initial population $\bf{S}$ $\leftarrow$ Set of
            \NoNumberAlgo{${\bf X}_{\rm uabs},{\bf S}_{\rm mbs}^{\rm ICIC},{\bf S}_{\rm pbs}^{\rm ICIC},{\bf S}_{\rm uabs}^{\rm ICIC}$}}
            \State \texttt{Evaluate Initial Population: $C_{\rm \mathbf{KPI}}(.)$}
             \State \texttt{Stop Condition $\leftarrow $ number of iterations}
             \While {\texttt{!Stop Condition}}
             \For{\texttt{$k=1:SZ_{\rm HM}$}}
                    \If{\texttt{$rand() < R_{\rm HMC}$}}
                        \State \texttt{Pick Best state, $\bf{S^\prime}$ from HM}
                        \If{\texttt{$rand() < par$}}
                            \State \texttt{Pitch adjustment on $\bf{S^\prime}$}
                            \State ${\bf S}_{rand}^{new} = {\bf S}_{rand}^\prime + (2 \times rand()-1)$ \\ \hspace{102pt} $\times fr_{rand}$
                        \Else 
                            \State \texttt{Crossover between ${\bf S}^k$ and \\ \hspace{57pt} a random member ${\bf S}^{rand}$}
                        \EndIf
                    \Else
                        \State \texttt{ Random selection ${\bf S}^{new}$}
                        \State ${\bf S}_{i}^{new} = (u_i - l_i) \times rand() + l_i$
                    \EndIf
                    \State \texttt{Evaluate Population: $C_{\rm \mathbf{KPI}}(.)$}    
                  \EndFor
                \State $fr \leftarrow fr \times 99\%$
            \EndWhile
            \State \texttt{Return ${\rm \mathbf{KPI}}$, Best state $\bf{S^\prime} \leftarrow$ Best solution}
        \EndProcedure
    \end{algorithmic}
    \vspace{-1mm}
\end{algorithm}

\subsection{Heuristic Algorithms}
With the purpose of improving computational efficiency and obtain the diverse optimal solution, we consider the GA and eHSGA as the heuristic algorithms~\cite{merwaday2016improved,kumbhar2018exploiting, binol2018hybrid} to simultaneously optimize UABS locations and ICIC parameters in the large search space. 

\textit{Genetic algorithm} considered in this article follows the approach similar to that in~\cite{merwaday2016improved,kumbhar2018exploiting}. This technique considers a population of candidate solutions which is evolved towards an optimal solution or near-optimal solution. Each candidate solution has a set of chromosomes that are evaluated and then altered and mutated to form next-generation offspring~\cite{binol2018time}. Through an iterative process,  adaptive-fit individuals in a population and environment are obtained. In this GA approach, the UABS coordinates (${\bf X}_{\rm uabs}$) and ICIC network parameters (${\bf S}_{\rm mbs}^{\rm ICIC},{\bf S}_{\rm pbs}^{\rm ICIC},{\bf S}_{\rm uabs}^{\rm ICIC}$) form the GA population, and a subsequent chromosome is illustrated in Fig.~\ref{GAChromosome}. With crossover rate of $cxr$ and mutation rate of $mr$ probabilities for a GA population size of $SZ_{\rm GM }$, the main steps used to optimize the UABS locations and ICIC network parameters while computing the maximum 5pSE and coverage probability is described in Algorithm~\ref{Alg:GaAlgo}. However, GA has limitations in terms of low convergence speed and requires high computation time.

Further, to obtain possible improvement over GA, we explore \textit{elitist harmony search based on the genetic algorithm} proposed in \cite{binol2018hybrid}.
We extend the approach of the proposed hybrid algorithm to optimize the UABS locations and ICIC network parameters. In the main procedure, the initial population generated using GA is considered as the harmony memory and the chromosome illustrated in Fig.~\ref{GAChromosome} as the harmony. Let ${\bf S}_{i}$ be the $i$th element of harmony ${\bf S}$; $u_i$ and $l_i$ are the upper/lower bounds of the $i$th variable; $rand()$ is a uniformly-distributed real random number in $[0,1]$~\cite{binol2018hybrid}. Then, we initialize the eHSGA parameters such as the harmony memory size ($SZ_{\rm HM}$), harmony memory consideration rate ($R_{\rm HMC}$), pitch adjustment rate ($par$), maximum number of improvisation ($N_{\rm IMP}$) and fret width ($fr$). The $R_{\rm HMC}$ and $par$ parameters in harmony search are critical to controlling the performance and speed of the convergence of the solution. To guarantee that the hybrid search method can expeditiously detect its way by avoiding local optima and the solution reached is diverse, $R_{\rm HMC}$ is updated linearly decreasing with the iteration and $par$ is dynamically adapted in linearly increasing rates. Then, we evaluate the fitness of every harmony in the harmony memory and sort the harmony memory in descending order of best fitness. This sorting ensures the harmony memory head always points to the best harmony member. Subsequently, using selection, crossover, and mutation, new harmony memory is generated. A merge rule is applied to sorted harmony memory and new harmony memory to generate an elitist harmony memory. As described in Algorithm~\ref{Alg:eHsgaAlgo}, through an iterative process, elitism is employed in the search process of obtaining optimal UABS locations and ICIC network parameters.

\begin{figure*}[!t]
\centering
\begin{subfigure}[b]{0.33\textwidth}
\label{fig:HexCovProb1}
\includegraphics[width=1.01\textwidth]{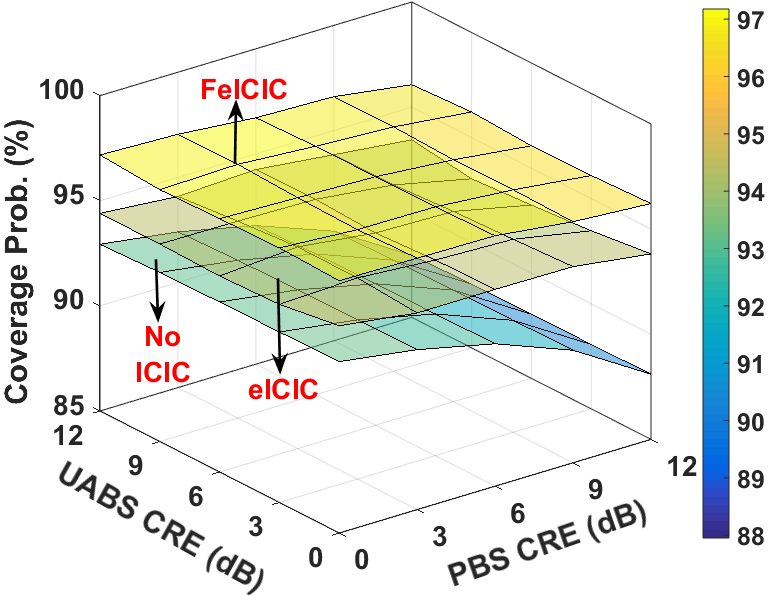}
\caption{Coverage probability vs. CRE.}
\end{subfigure}
\begin{subfigure}[b]{0.33\textwidth}
\label{fig:Hex5pSE1}
\includegraphics[width=1.06\textwidth]{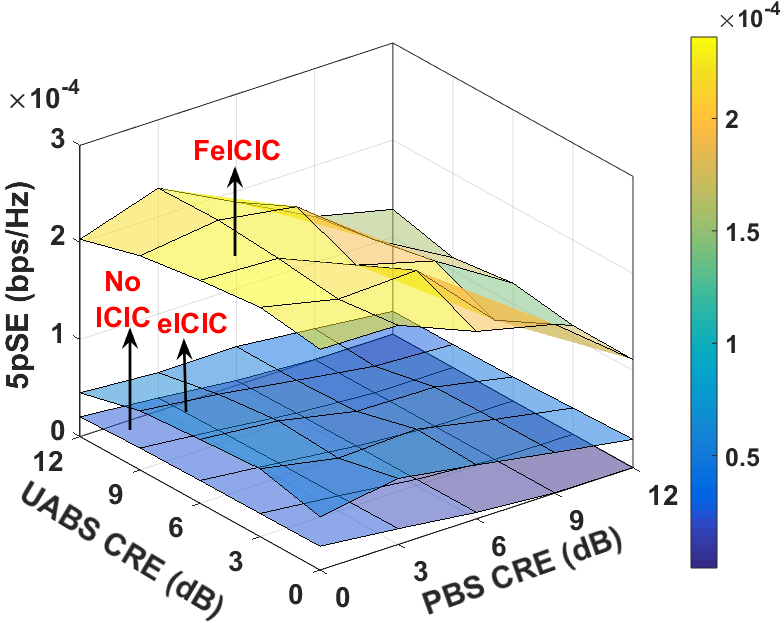}
\caption{Peak 5pSE vs. CRE.}
\end{subfigure}
\begin{subfigure}[b]{0.33\textwidth}
\label{fig:Hex5pSE2}
\includegraphics[width=1\textwidth]{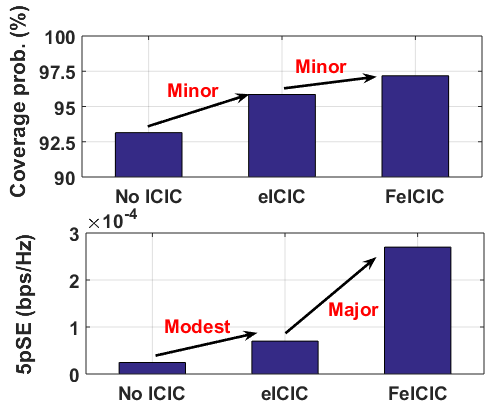}
\caption{Performance improvement of the two KPIs.}
\end{subfigure}
\caption{The effects of combined CRE at PBS and UABS on the two KPIs of the network, with and without ICIC; when the UABS are deployed at height of 25 m.}
\label{fig:HexKpiH25}
\end{figure*}

\begin{table}[!t]
\processtable{System and simulation parameters.\label{tab:SysParams}}
{\begin{tabular*}{20pc}{@{\extracolsep{\fill}}ll@{}}\toprule
Parameter  & Value \\
\midrule
Simulation area ($A_{\rm sim}$)  & $100 {\rm\ km^2}$\\ 
MBS, PBS, GUE, AUE densities & 4, 12, 100, and 1.8 per km$^2$ \\
Number of UABS & 60 \\ 
MBS, PBS, and UABS transmit powers & 46, 30, and 26 dBm\\ 
Height of MBS, PBS, and UABS & 36 and 15m\\ 
Height of UABS & 25, 36, and 50 m\\ 
Height of GUE and AUE & 1.5 and 22.5 m\\ 
PSC LTE Band~14 center frequency & 763 MHz for downlink\\ 
Power reduction factor $\alpha_{\rm mbs}$ and $\alpha_{\rm pbs}$ &  $0$ to $1$  \\
USF Duty cycle $\beta_{\rm mbs}$, $\beta_{\rm pbs}$ &  $0$ to $100$\%   \\ 
Scheduling threshold for MUEs ($\rho_{\rm mbs}$)            &  $20$ dB to $40$ dB \\ 
Scheduling threshold for PUEs ($\rho_{\rm pbs}$)            &  $-10$ dB to $10$ dB \\ 
Scheduling threshold for UUEs ($\rho_{\rm uabs}$)      &  $-5$ dB to $5$ dB \\ 
Range expansion bias for $\tau_{\rm uabs}$, $\tau_{\rm uabs}$ & $0$ dB to $12$ dB \\ 
GA population size ($SZ_{\rm GA}$) and generation number & $60$ and $100$\\
GA crossover($cxr$) and mutation ($mr$) probabilities & $0.7$ and $0.1$\\
eHSGA population size ($SZ_{\rm HM}$) & 60 \\
Harmonic memory pitch adjustment rate ($par$)  & $\max=0.8$, $\min=0.4$ \\
Harmonic memory consideration rate ($R_{\rm HMC}$) & $\max=0.8$, $\min=0.2$\\
Harmonic memory fret ( $fr$) &  $1$\\
\botrule
\end{tabular*}}{}
\vspace{-2mm}
\end{table}

\section{Simulation Results}
\label{simRes}
In this section, with the help of extensive Matlab-based computer simulation and system parameters set to the values given in Table~\ref{tab:SysParams}, we conduct a comparative study of the two KPIs of the proposed AG-HetNet, with/without ICIC techniques for different deployment heights of UABS and while considering brute-force, GA, and eHSGA optimization techniques. In order to reduce the complexity of the optimization algorithms and simulation runtime, the deployment height of UABS is not considered when optimizing UABS locations, i.e., the UABS locations are optimized in 2D. However, in this article, we do investigate and compare the performance of the KPIs by manually deploying UABS at practical heights of $25$ m, $36$ m, and $50$ m.

\subsection{KPI Optimization using Brute Force Technique}
\begin{figure*}[!t]
\centering
\begin{subfigure}[b]{0.33\textwidth}
\label{fig:HexCovProb2}
\includegraphics[width=1\textwidth]{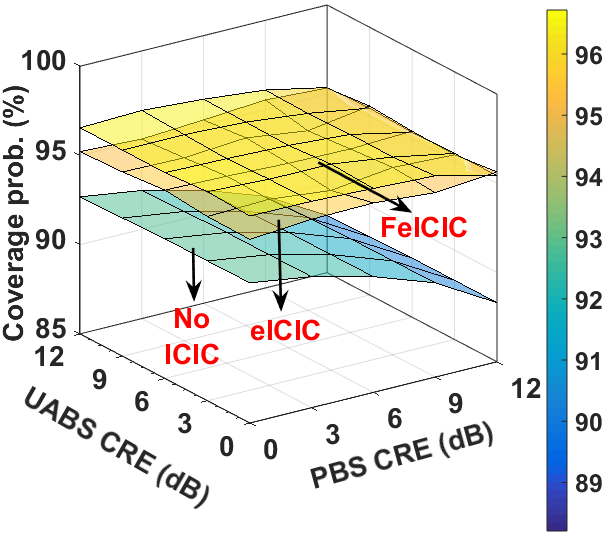}
\caption{Coverage probability vs. CRE.}
\end{subfigure}
\begin{subfigure}[b]{0.33\textwidth}
\label{fig:Hex5pSE3}
\includegraphics[width=1.025\textwidth]{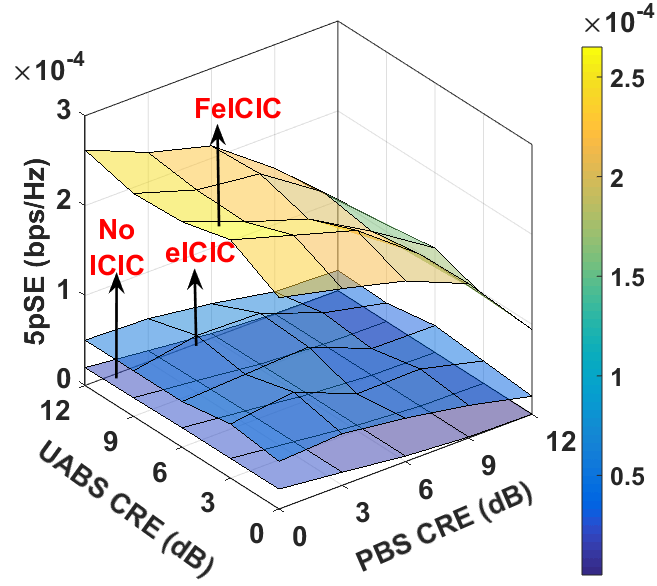}
\caption{Peak 5pSE vs. CRE.}
\end{subfigure}
\begin{subfigure}[b]{0.33\textwidth}
\label{fig:Hex5pSE4}
\includegraphics[width=1.02\textwidth]{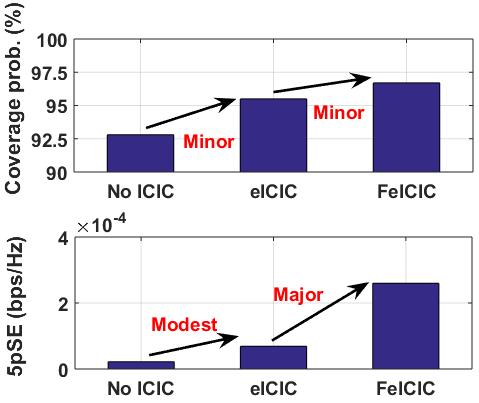}
\caption{Performance improvement of the two KPIs.}
\end{subfigure}
\caption{The effects of combined CRE at PBS and UABS on the two KPIs of the network, with and without ICIC; when UABS are deployed at height of 36 m.}
\label{fig:HexKpiH36} 
\end{figure*}

\begin{figure*}[!t]
\centering
\begin{subfigure}[b]{0.33\textwidth}
\label{fig:HexCovProb3}
\includegraphics[width=1\textwidth]{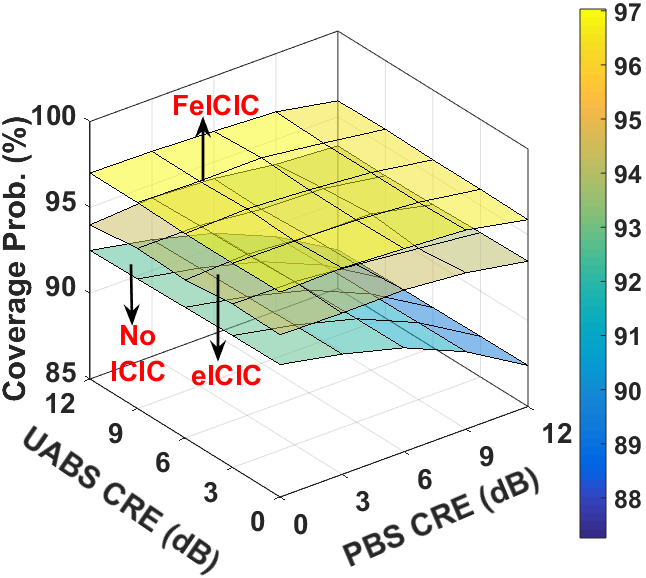}
\caption{Coverage prob. vs. CRE.}
\end{subfigure}
\begin{subfigure}[b]{0.33\textwidth}
\label{fig:Hex5pSE6}
\includegraphics[width=1.045\textwidth]{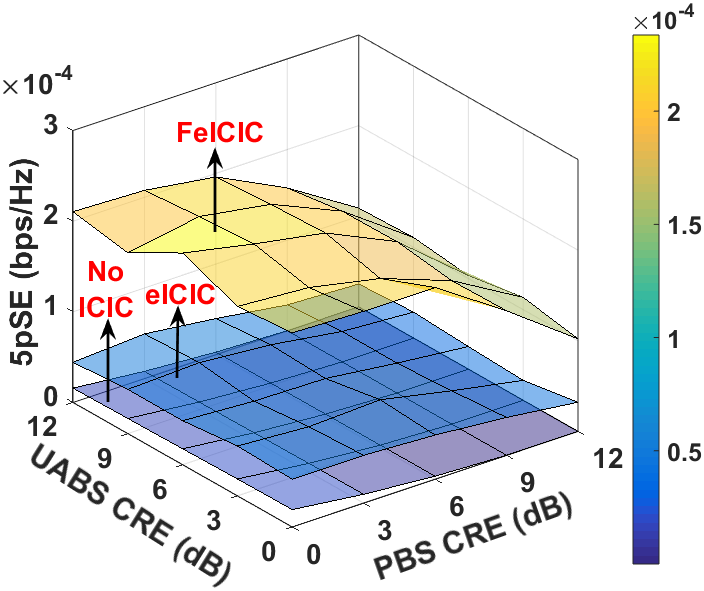}
\caption{Peak 5pSE vs. CRE.}
\end{subfigure}
\begin{subfigure}[b]{0.33\textwidth}
\label{fig:Hex5pSE7}
\includegraphics[width=0.95\textwidth]{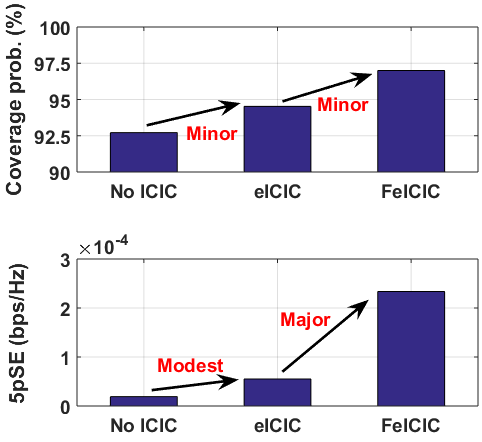}
\caption{Performance improvements of the two KPIs.}
\end{subfigure}
\caption{The effects of combined CRE at PBS and UABS on the two KPIs of the network, with and without ICIC; when UABS are deployed at height of 50 m.}
\label{fig:HexKpiH50}
\end{figure*}

The 3D surface plot in Fig.~\ref{fig:HexKpiH25}, Fig.~\ref{fig:HexKpiH36}, and Fig.~\ref{fig:HexKpiH50} illustrates the combined effect of CRE at PBSs and UABSs (along x- and y-axes) on the coverage probability and 5pSE (along the z-axis) of the wireless network. In an initial inspection of Fig.~\ref{fig:HexKpiH25}, Fig.~\ref{fig:HexKpiH36} and Fig.~\ref{fig:HexKpiH50}, we can intuitively conclude that FeICIC performs better when compared to eICIC and without any ICIC techniques. The comparative analysis of Fig.~\ref{fig:HexKpiH25}(c), Fig.~\ref{fig:HexKpiH36}(c), and Fig.~\ref{fig:HexKpiH50}(c) reveals that the improvement in coverage probability is less significant, but the 5pSE improvement is significant.

\begin{figure*}[!t]
\centering
\begin{subfigure}[b]{0.33\textwidth}
\label{fig:GaKpiH25}
\includegraphics[width=1\textwidth]{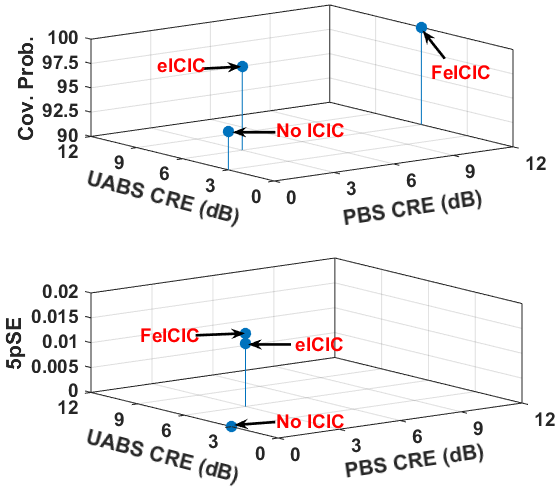}
\caption{Peak value observation when UABS are deployed at the height of 25 m.}
\end{subfigure} 
\begin{subfigure}[b]{0.33\textwidth}
\label{fig:GaKpiH36}
\includegraphics[width=1\textwidth]{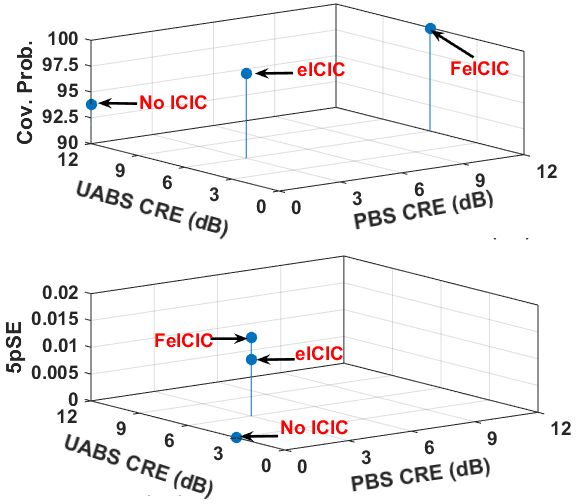}
\caption{Peak value observation when UABS are deployed at the height of 36 m.}
\end{subfigure}
\begin{subfigure}[b]{0.33\textwidth}
\label{fig:GaKpiH50}
\includegraphics[width=1\textwidth]{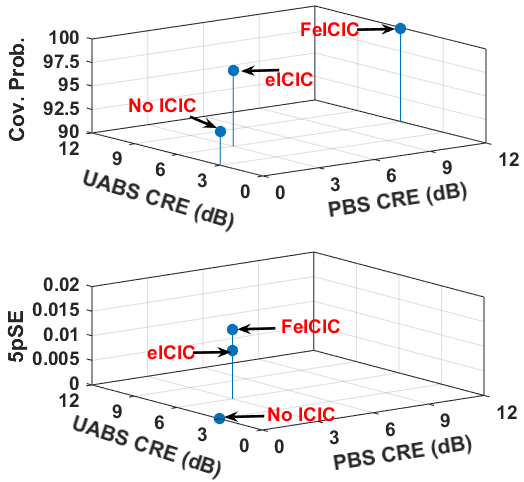}
\caption{Peak value observation when UABS are deployed at the height of 50 m.}
\end{subfigure}
\begin{subfigure}[b]{0.33\textwidth}
\label{fig:GaPerCompareH25}
\includegraphics[width=1\textwidth]{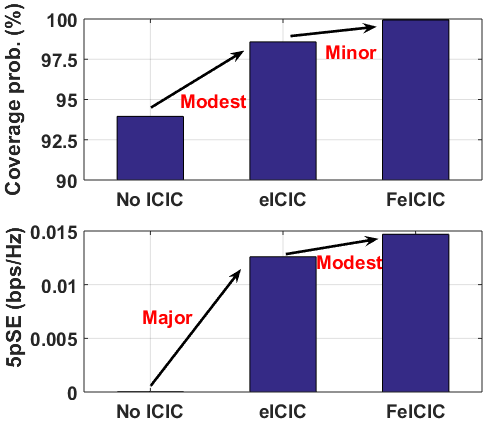}
\caption{Performance improvement when UABS are deployed at the height of 25 m.}
\end{subfigure}
\begin{subfigure}[b]{0.33\textwidth}
\label{fig:GaPerCompareH36}
\includegraphics[width=1\textwidth]{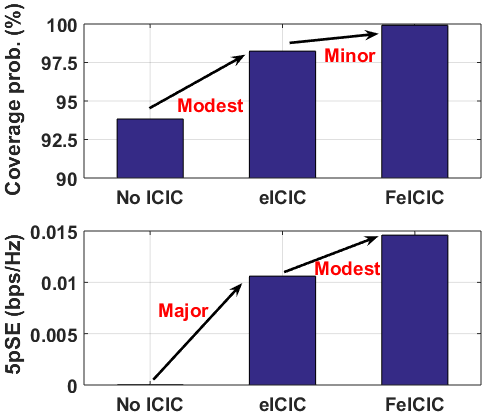}
\caption{Performance improvement when UABS are deployed at the height of 36 m.}
\end{subfigure}
\begin{subfigure}[b]{0.33\textwidth}
\label{fig:GaPerCompareH50}
\includegraphics[width=1\textwidth]{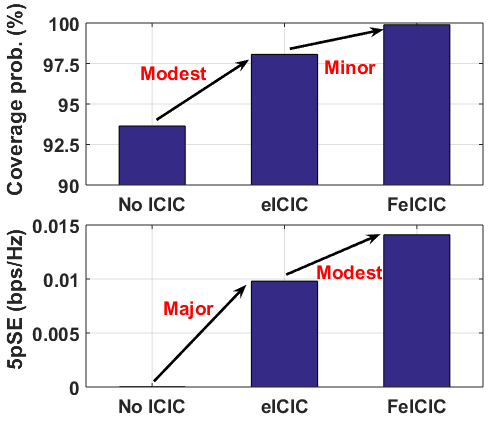}
\caption{Performance improvements when UABS are deployed at the height of 50 m.}
\end{subfigure}
\caption{A combined effect of CRE at PBS and UABS on the two KPIs of the network, with and without ICIC. When UABS are deployed at the height of $25$ m, $36$ m, and $50$ m and while considering GA.}
\label{fig:GaPerf} 
\vspace{-4mm}
\end{figure*}

 When UABS is deployed at a height higher than PBS but lower than MBS, i.e., UABS deployment height is $25$m, coverage probability with eICIC sees a minor improvement over the absence of any ICIC, and FeICIC also sees a minor improvement over eICIC. With increasing CRE of UABS and lower CRE for PBS, the peak values of the coverage probability for the ICIC techniques are observed when the UABS CRE is between $9-12$ dB, and PBS CRE varies between $0-3$ dB. For 5pSE, eICIC sees modest improvement over the absence of any ICIC, and FeICIC sees a major improvement over eICIC. The peak values of 5pSE for the ICIC techniques are observed for lower values of CRE between $3-6$ dB for both UABS and PBS. 

Whereas, when UABS are deployed at the same height as MBS, i.e., UABS deployment height is $36$m, coverage probability with eICIC sees a minor improvement over the absence of any ICIC, and FeICIC also sees a minor improvement over eICIC. The peak values of coverage probability for the ICIC techniques are observed for lower values of CRE between $3-6$ dB for both UABS and PBS. For 5pSE, eICIC sees modest improvement over the absence of any ICIC, and FeICIC sees a major improvement of over eICIC. With increasing CRE of UABS and lower CRE for PBS, the peak values of the 5pSE for the ICIC techniques are observed when the UABS CRE is between $9-12$ dB, and PBS CRE varies between $0-3$ dB. 

Finally, when the UABS is deployed at a height higher than MBS, i.e., UABS deployment height is $50$m, coverage probability with eICIC sees a minor improvement over the absence of any ICIC, and FeICIC also sees a minor improvement over eICIC. With increasing CRE of UABS and lower CRE for PBS, the peak values of the coverage probability for the ICIC techniques are observed when the UABS CRE is between $9-12$ dB, and PBS CRE varies between $0-3$ dB. For 5pSE, eICIC sees modest improvement over the absence of any ICIC, and FeICIC sees a major improvement of over eICIC. With increasing CRE of UABS and lower CRE for PBS, the peak values of the 5pSE for the ICIC techniques are observed when the UABS CRE is $9$ dB, and PBS CRE varies between $0-3$ dB.

Overall, when UABSs are deployed on a fixed hexagonal grid and using the brute-force technique to optimize ICIC parameters, the peak values of 5pSE and coverage probability is observed when UABS is deployed at the low altitude of $25$ m and using Rel-11 reduced power FeICIC technique as seen in Fig.~\ref{fig:HexKpiH25}, Fig.~\ref{fig:HexKpiH36}, and Fig.~\ref{fig:HexKpiH50}.

\begin{figure*}[!t]
\centering
\begin{subfigure}[b]{0.33\textwidth}
\label{fig:ehsgaKpiH25}
\includegraphics[width=1\textwidth]{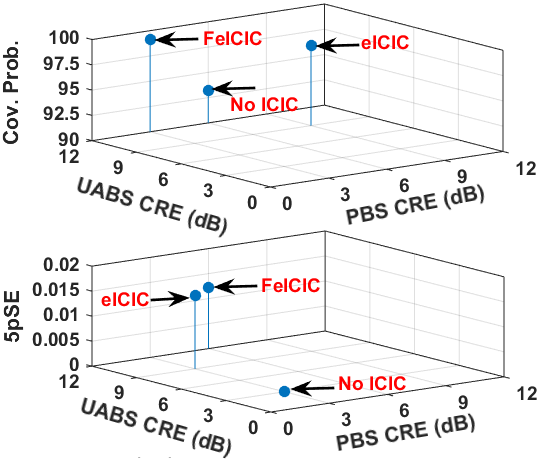}
\caption{Peak value observation when UABS are deployed at the height of 25m.}
\end{subfigure} 
\begin{subfigure}[b]{0.33\textwidth}
\label{fig:ehsgaKpiH36}
\includegraphics[width=1\textwidth]{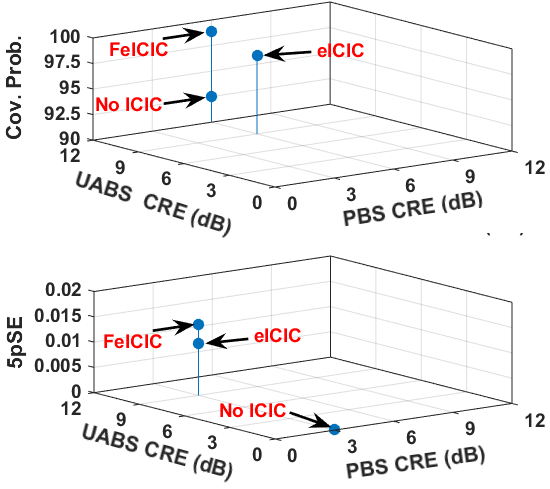}
\caption{Peak value observation when UABS are deployed at the height of 36m.}
\end{subfigure}
\begin{subfigure}[b]{0.33\textwidth}
\label{fig:ehsgaKpiH50}
\includegraphics[width=1\textwidth]{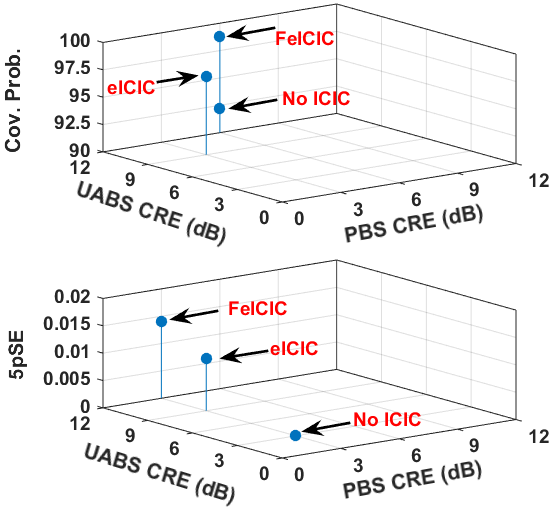}
\caption{Peak value observation when UABS are deployed at the height of 50m.}
\end{subfigure}
\begin{subfigure}[b]{0.33\textwidth}
\label{fig:ehsgaPerCompareH25}
\includegraphics[width=1\textwidth]{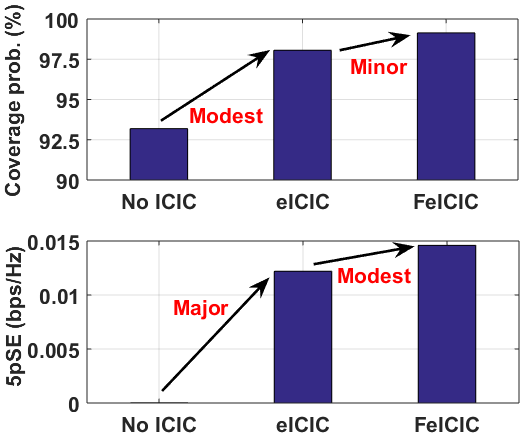}
\caption{Performance improvement when UABS are deployed at the height of 25m.}
\end{subfigure}
\begin{subfigure}[b]{0.33\textwidth}
\label{fig:ehsgaPerCompareH36}
\includegraphics[width=1\textwidth]{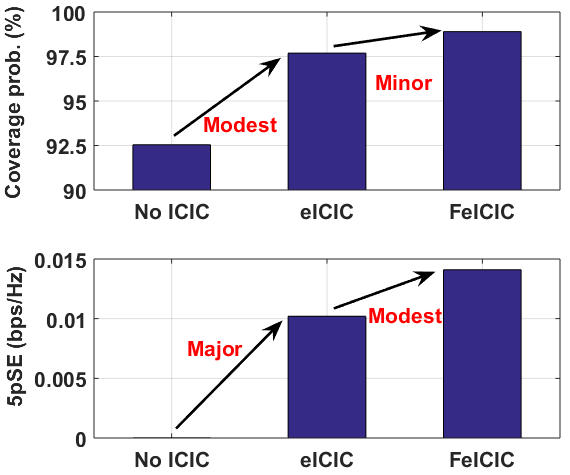}
\caption{Performance improvement when UABS are deployed at the height of 36m.}
\end{subfigure}
\begin{subfigure}[b]{0.33\textwidth}
\label{fig:ehsgaPerCompareH50}
\includegraphics[width=1\textwidth]{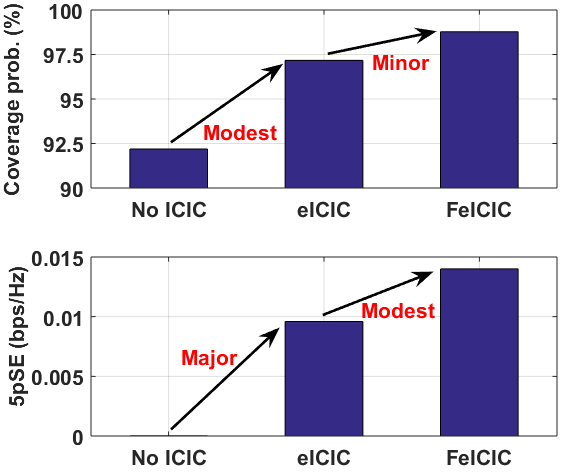}
\caption{Performance improvements when UABS are deployed at the height of 50m.}
\end{subfigure}
\caption{A combined effect of CRE at PBS and UABS on the two KPIs of the network, with and without ICIC. When UABS are deployed at the height of $25$ m, $36$ m, and $50$ m and while considering eHSGA.}
\label{fig:ehsgaPerf} 
\end{figure*}

\subsection{KPI Optimization using Genetic Algorithm}
In the following, we discuss the key observations when UABS locations and ICIC network parameters are jointly optimized through GA. In Fig.~\ref{fig:GaPerf} we plot the peak coverage probability and 5pSE values (along the z-axis) against the combined effect of CRE at PBSs and UABSs (along x- and y-axes) of the wireless network, when UABS are deployed at the height of $25$ m, $36$ m, and $50$ m and while considering GA. In an initial inspection of Fig.~\ref{fig:GaPerf}(a), Fig.~\ref{fig:GaPerf}(b), and Fig.~\ref{fig:GaPerf}(c), we can intuitively conclude that FeICIC performs better when compared to eICIC and without any ICIC techniques. 

We present the comparative analysis of the peak value observations of the two KPIs in Fig.~\ref{fig:GaPerf}(d), Fig.~\ref{fig:GaPerf}(e), and Fig.~\ref{fig:GaPerf}(f). When UABS is deployed at a height higher than PBS but lower than MBS, i.e., UABS deployment height is $25$ m, coverage probability with eICIC sees modest improvement in the absence of any ICIC, and FeICIC sees a minor improvement over eICIC. The peak values of coverage probability for the ICIC techniques are observed for CRE values between $3-6$ dB for UABS, and PBS CRE varies between $0-12$ dB. For 5pSE, eICIC sees a major improvement in the absence of any ICIC, and FeICIC sees a modest improvement over eICIC. The peak values of 5pSE for the ICIC techniques are observed for moderate values of CRE between $3-6$ dB for UABS and is $0-3$ dB for PBS. 

Whereas, when UABS are deployed at the same height as MBS, i.e., UABS deployment height is $36$ m, coverage probability with eICIC sees modest improvement in the absence of any ICIC, and FeICIC sees a minor improvement over eICIC. The peak values of coverage probability for the ICIC techniques are observed for larger values of CRE between $6-12$ dB for UABS, and PBS CRE varies between $0-12$ dB. For 5pSE, eICIC sees a major improvement over the absence of any ICIC, and FeICIC sees a minor improvement over eICIC. The peak values of 5pSE for the ICIC techniques are observed for moderate values of CRE between $3-6$ dB for UABS and is $0-3$ dB for PBS. 

Finally, when the UABS is deployed at a height higher than MBS, i.e., UABS deployment height is $50$ m, coverage probability with eICIC sees modest improvement over the absence of any ICIC, and FeICIC sees a minor improvement over eICIC. The peak values of coverage probability for the ICIC techniques are observed for moderate values of CRE between $3-6$ dB for UABS, and PBS CRE varies between $0-12$ dB. For 5pSE, eICIC sees a major improvement over the absence of any ICIC, and FeICIC sees a minor improvement over eICIC. The peak values of 5pSE for the ICIC techniques are observed for moderate values of CRE between $3-6$ dB for UABS and is $0-3$ dB for PBS.

Overall, using GA to joint optimize the UABS locations and ICIC parameters, the peak values of 5pSE and coverage probability are observed when UABS is deployed at a low altitude of $25$ m and using Rel-11 reduced power FeICIC technique as seen in Fig.~\ref{fig:GaPerf}. 

\subsection{KPI optimization using eHSGA}
In the following, we discuss the key observations when UABS locations and ICIC network parameters are jointly optimized through the eHSGA algorithm. In Fig.~\ref{fig:ehsgaPerf} we plot the peak coverage probability and 5pSE values (along the z-axis) against the combined effect of CRE at PBSs and UABSs (along x- and y-axes) of the wireless network, when UABS are deployed at the height of $25$ m, $36$ m, and $50$ m. In an initial inspection of Fig.~\ref{fig:ehsgaPerf}(a), Fig.~\ref{fig:ehsgaPerf}(b), and Fig.~\ref{fig:ehsgaPerf}(c), we can intuitively conclude that FeICIC performs better when compared to eICIC and without any ICIC techniques. 

Whereas, in Fig.~\ref{fig:ehsgaPerf}(d), Fig.~\ref{fig:ehsgaPerf}(e), and Fig.~\ref{fig:ehsgaPerf}(f) we compare the peak value of the two KPIs. When UABS is deployed at a height higher than PBS but lower than MBS, i.e., UABS deployment height is $25$ m, coverage probability with eICIC sees modest improvement over the absence of any ICIC, and FeICIC sees a minor improvement over eICIC. The peak values of coverage probability for the ICIC techniques are observed for higher values of CRE between $9-12$ dB for UABS and moderate CRE values between $3-9$ dB for PBS. For 5pSE, eICIC sees a major improvement over the absence of any ICIC, and FeICIC sees a modest improvement over eICIC. The peak values of 5pSE for the ICIC techniques observed for the CRE values between $3-12$ dB for UABS and lower CRE values between $3-6$ dB for PBS.

Whereas, when UABS are deployed at the same height as MBS, i.e., UABS deployment height is $36$ m, coverage probability with eICIC sees modest improvement in the absence of any ICIC, and FeICIC sees a minor improvement over eICIC. The peak values of coverage probability for the ICIC techniques are observed for higher values of CRE between $9-12$ dB for UABS and moderate CRE values between $3-6$ dB for PBS. For 5pSE, eICIC sees a major improvement over the absence of any ICIC, and FeICIC sees a minor improvement over eICIC. The peak values of 5pSE for the ICIC techniques are observed for CRE values between $0-9$ dB for UABS and is $3$ dB for PBS. 

Finally, when the UABS is deployed at a height higher than MBS, i.e., UABS deployment height is $50$m, coverage probability with eICIC sees modest improvement in the absence of any ICIC, and FeICIC sees a minor improvement over eICIC. The peak values of coverage probability for the ICIC techniques are observed for higher values of CRE between $9-12$ dB for UABS and moderate CRE values between $3-6$ dB for PBS. For 5pSE, eICIC sees a major improvement in the absence of any ICIC, and FeICIC sees a minor improvement over eICIC. The peak values of 5pSE for the ICIC techniques are observed for CRE values between $3-12$ dB for UABS and is $3$ dB for PBS.

Overall, using eHSGA to joint optimize the UABS location and ICIC parameters, the peak values of 5pSE and coverage probability are higher when UABS is deployed at a low altitude of $25$ m and using Rel-11 reduced power FeICIC technique as seen in Fig.~\ref{fig:ehsgaPerf}. 

\begin{table*}[!t]
\caption{Coverage probability peak value observations in \%.}
\centering
\begin{tabular}{l ccc ccc ccc}
\toprule
 & \multicolumn{3}{c}{\textbf{Brute force}} & \multicolumn{3}{c}{\textbf{Genetic algorithm}} & \multicolumn{3}{c}{\textbf{eHSGA}}\\
 \cmidrule(lr){2-4} \cmidrule(lr){5-7} \cmidrule(lr){8-10}
 & \multicolumn{3}{c}{\textbf{UABS height}} & \multicolumn{3}{c}{\textbf{UABS height}} & \multicolumn{3}{c}{\textbf{UABS height}} \\
\cmidrule(lr){2-4} \cmidrule(lr){5-7} \cmidrule(lr){8-10}
ICIC     & \textbf{25m}   & \textbf{36m}  & \textbf{50m} & \textbf{25m}   & \textbf{36m}  & \textbf{50m} & \textbf{25m}   & \textbf{36m}  & \textbf{50m}\\
\midrule
No ICIC  & $93.15$     & $92.86$      & $92.71$      & $93.95$       & $93.83$      & $93.64$  & $93.19$ & $92.54$ & $92.19$\\
eICIC    & $95.85$     & $95.62$      & $94.52$      & $98.58$       & $98.24$      & $98.06$  & $97.89$ & $97.69$ & $97.17$\\
FeICIC   & $97.18$     & $96.99$      & $96.72$      & $99.94$       & $99.92$      & $99.89$  & $99.14$ & $98.90$ & $98.78$\\
\bottomrule
\end{tabular}
\label{tab:CovProbPeakVal}
\end{table*}

\begin{table*}[!t]
\caption{5pSE peak value observations in bps/kHz.}
\centering
\begin{tabular}{l ccc ccc ccc}
\toprule
 & \multicolumn{3}{c}{\textbf{Brute force}} & \multicolumn{3}{c}{\textbf{Genetic algorithm}} & \multicolumn{3}{c}{\textbf{eHSGA}}\\
 \cmidrule(lr){2-4} \cmidrule(lr){5-7} \cmidrule(lr){8-10}
 & \multicolumn{3}{c}{\textbf{UABS height}} & \multicolumn{3}{c}{\textbf{UABS height}} & \multicolumn{3}{c}{\textbf{UABS height}} \\
\cmidrule(lr){2-4} \cmidrule(lr){5-7} \cmidrule(lr){8-10}
ICIC     & \textbf{25m}   & \textbf{36m}  & \textbf{50m} & \textbf{25m}   & \textbf{36m}  & \textbf{50m} & \textbf{25m}   & \textbf{36m}  & \textbf{50m}\\
\midrule
No ICIC       & $2.45e-5$     & $2.21e-5$      & $1.92e-5$      & $4.01e-5$       & $3.53e-5$      & $3.44e-5$  & $2.63e-5$ & $2.30e-5$ & $1.98e-5$\\
eICIC & $0.70e-4$     & $0.69e-4$      & $0.55e-4$      & $1.26e-2$       & $1.06e-2$      & $0.98e-2$ & $1.22e-2$ & $1.02e-2$ & $0.96e-2$\\
FeICIC   & $0.27e-3$     & $0.24e-3$      & $0.23e-3$      & $1.48e-2$       & $1.46e-2$      & $1.41e-2$  & $1.46e-2$ & $1.41e-2$ & $1.40e-2$\\
\bottomrule
\end{tabular}
\label{tab:5psePeakVal}
\end{table*}

\subsection{Performance Comparison of the Algorithms}
We summarize our key results from earlier simulations and compare the computation time when using brute-force, GA, and eHSGA techniques with/without ICIC optimization. 

\subsubsection{Comparison of KPIs}
From the simulation results given in Fig.~\ref{fig:HexKpiH25}, Fig.~\ref{fig:HexKpiH36}, Fig.~\ref{fig:HexKpiH50}, Fig.~\ref{fig:GaPerf}, and Fig.~\ref{fig:ehsgaPerf},  we observe reduced power FeICIC in Rel-11 is seen to outperform Rel-10 eICIC and without ICIC in terms of the overall 5pSE and coverage probability of the AG-HetNet. Further inspection reveals that the heuristic techniques (GA and eHSGA) outperform the brute-force technique and observe significant improvement in terms of overall 5pSE and coverage probability of the AG-HetNet. In particular, GA meta-heuristic technique achieved a marginal 5pSE and coverage probability gains of upto $3$\% over the hybrid eHSGA optimization technique. 

We also observe that the performance of the wireless network is optimal when UABSs are deployed at a height of $25$ m. However, as the UABS deployment height increases to $36$ m and $50$ m, a gradual decrease in the 5pSE and coverage probability of the wireless network is observed. The higher deployment heights of the UABS improve LOS to the UEs but also increases interference probability with UEs in cell-edge/CRE, thus degrading the overall performance of the AG-HetNet. 
 
 We summarize the peak values observed for coverage probability and 5pSE for with/without ICIC techniques for different deployment heights of UABS; while using brute force, genetic algorithm, and elitist hybridization between harmonic search and genetic algorithm in Table~\ref{tab:5psePeakVal} and Table~\ref{tab:CovProbPeakVal}.
 
\begin{figure} [!t]
\centering
\includegraphics[width=0.9\linewidth]{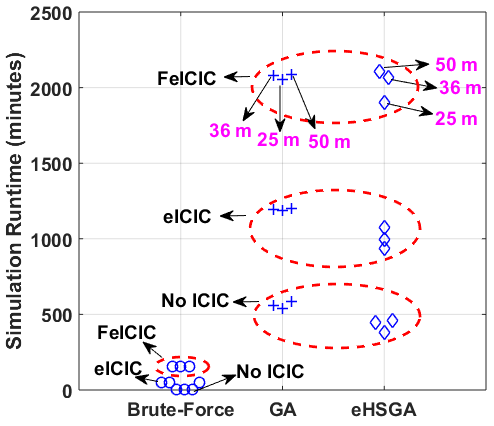}
\caption{Simulation runtime for evaluating a single KPI with/without ICIC techniques; when UABSs are deployed at different heights and using different optimization techniques.}
\label{simulationRuntime}
\end{figure}

\subsubsection{Comparison of Computational Complexity Gains}
In this subsection, we compare the computational complexity gains for brute-force, GA, and eHSGA techniques with/without ICIC optimization and when UABSs are deployed at $25$ m, $36$ m, and $50$ m. Using the NCSU high-performance computing server, and Monte-Carlo experimental approach, we calculate the mean runtime for Matlab simulation while evaluating an individual KPI. In Fig.~\ref{simulationRuntime}, we plot the mean runtime required for calculating the peak KPI value with optimal ICIC network parameters and UABS locations using~\eqref{Eq:optimizeState}, brute-force given in Algorithm~\ref{Alg:bruteForceAlgo}, GA given in Algorithm~\ref{Alg:GaAlgo}, eHSGA given in Algorithm~\ref{Alg:eHsgaAlgo}, and the simulation values defined in Table~\ref{tab:SysParams}.

In an initial inspection of Fig.~\ref{simulationRuntime}, we observe reduced power FeICIC technique requires significantly higher computational time when compared to the ABS eICIC and without ICIC techniques, for different deployment heights of UABS ($25$ m, $36$ m, and $50$ m) and optimization techniques (brute-force, GA, and eHSGA). The reduced power FeICIC technique requires significantly higher computational time because the optimization of power reduction parameters ($\alpha_{\rm mbs}$ and $\alpha_{\rm pbs}$) of the three-tier LTE subframes, increases the scope of search space when compare to ABS eICIC and without ICIC techniques. Further analysis of Fig.~\ref{simulationRuntime} reveals that optimization (ICIC parameters and UABS locations) using GA and eHSGA both require significantly more computational time when compared to the brute-force technique. In particular, for lower complexity eICIC and without ICIC techniques, eHSGA observes substantial computational complexity gains between $10.65-29.14$\% over GA. Whereas with higher complexity reduced power FeICIC technique, eHSGA observes marginal computational complexity gains of upto $7$\% over GA.

Whereas, the UABS deployment height of $50$ m observes a sparse increment in the computation time when compared to $36$ m and $25$ m, while jointly optimizing the ICIC parameters and UABS locations in the same search space. The higher deployment heights of the UABS improve LOS to the UEs but also increases interference probability with UEs in cell-edge/CRE and consequently increases the computation time to optimize the ICIC parameters and UABS locations.

To summarize, although the computational complexity required to optimize the ICIC network parameters and UAV locations using heuristic techniques (GA and eHSGA) is higher, but is effective in achieving broadband rates. In particular, the heuristic techniques can meet the public safety network requirement of $95\%$ geographical coverage with broadband rates~\cite{moore2014first}. Furthermore, from Fig.~\ref{simulationRuntime}, Table~\ref{tab:CovProbPeakVal}, and Table~\ref{tab:5psePeakVal}, we observe hybrid eHSGA achieves marginal computational complexity gain over meta-heuristic GA technique. Whereas, optimization using GA marginally improves KPI gains when compared to optimization using eHSGA. Hence the determination of an appropriate heuristic algorithm, which achieves the trade-off between computational complexity and finding optimum or close to the optimum solution of a search problem in the real world, requires further investigation.

\section{Conclusion}
\label{conclusion}
In this paper, we provide system-level insights into the LTE-Advanced AG-HetNet and evaluate the network performance in terms of coverage probability and 5pSE. In particular, we integrate low-altitude unmanned vehicles as both AUE and UABS into an existing terrestrial network. While considering key design parameters such as the base-station heights, antenna 3DBF, path loss model specific to UE camping, interference coordination, and altitude variation of unmanned vehicles. Using these design considerations and through Monte-Carlo simulations, we maximized the coverage probability and 5pSE of the overall network, while mitigating intercell interference and optimizing ICIC parameters and UABS locations using brute-force and heuristics approach.

Finally, our analysis shows that the AG-HetNet with reduced power subframes (FeICIC) yields better coverage probability and 5pSE than that with almost blank subframes (eICIC) and without any ICIC. Our simulations results show that the heuristic algorithms (GA and eHSGA) outperform the brute-force technique and achieve effective peak values of coverage probability and 5pSE. In particular, optimization of higher complexity FeICIC using the GA technique achieves marginally better peak KPI values but requires slightly more computational time when compared to hybrid eHSGA. Although the trade-off exists between KPI gains and computational complex gains, simulation results show that hybrid eHSGA can be feasible and effective. We also found that the wireless networks performed sparely better when UABSs are deployed at the height of $25$ m compared to $36$ m and $50$ m deployment height.

\section{Acknowledgments}\label{sec11}
This research was supported in part by NSF under CNS-1453678. A portion of the
results in this manuscript has also been published in the IEEE Radio and Wireless Symposium (RWS) in 2019 \cite{kumbhar2018interference}.

\bibliographystyle{iet}
\bibliography{myBib.bib}

\begin{thebibliography}{10}

\bibitem{R1}
Kumbhar, A., Koohifar, F., Guvenc, I., Mueller, B.: `A survey on legacy and
  emerging technologies for public safety communications', \emph{IEEE Commun
  Survery Tuts},  2016, \textbf{18}, pp.~97--124

\bibitem{chandrasekharan2016designing}
Chandrasekharan, S., Gomez, K., Al.Hourani, A., Kandeepan, S., Rasheed, T.,
  Goratti, L., et~al.: `Designing and implementing future aerial communication
  networks', \emph{IEEE Commun Mag},  2016, \textbf{54}, (5), pp.~26--34

\bibitem{merwaday2016improved}
Merwaday, A., Tuncer, A., Kumbhar, A., Guvenc, I.: `{Improved Throughput
  Coverage in Natural Disasters: Unmanned Aerial Base Stations for
  Public-Safety Communications}', \emph{IEEE Vehic Technol Mag},  2016,
  \textbf{11}, (4), pp.~53--60

\bibitem{kumbhar2018exploiting}
Kumbhar, A., G{\"u}ven{\c{c}}, I., Singh, S., Tuncer, A.: `{Exploiting
  LTE-Advanced HetNets and FeICIC for UAV-assisted public safety
  communications}', \emph{IEEE Access},  2018, \textbf{6}, pp.~783--796

\bibitem{attCow}
{AT\&T}.
\newblock `{Flying COW Connects Puerto Rico}'.
\newblock (,  2017.
\newblock Available from:
  \url{https://about.att.com/inside_connections_blog/flying_cow_puertori}

\bibitem{drive1}
{The Drive}.
\newblock `{AT\&T and Verizon Test 4G LTE Drones in New Jersey}'.
\newblock (,  2018.
\newblock Available from:
  \url{https://www.thedrive.com/tech/21756/att-and-verizon-test-4g-lte-drones-in-new-jersey}

\bibitem{cnbc1}
{CNBC}.
\newblock `{AT\&T and Verizon drones provide cell service in natural
  disasters}'.
\newblock (,  2018.
\newblock Available from:
  \url{https://www.cnbc.com/2018/06/22/att-and-verizon-drones-provide-cell-service-in-natural-disasters.html}

\bibitem{VZWSmallCell}
Gibbs, C.: `{Verizon claims largest small cell deployment in the U.S}',
  \emph{{Fierce Wireless}},  2017, Available from:
  \url{http://www.fiercewireless.com/wireless/verizon-claims-largest-small-cell-deployment-any-u-s-carrier}

\bibitem{ATTSmallCell}
Heisler, Y.: `{AT\&T wants to use drones to improve its LTE network}',
  \emph{{Yahoo Tech News}},  2016, Available from:
  \url{https://www.yahoo.com/tech/t-wants-drones-improve-lte-network-193203323.html}

\bibitem{nakamura2013trends}
Nakamura, T., Nagata, S., Benjebbour, A., Kishiyama, Y., Hai, T., Xiaodong, S.,
  et~al.: `{Trends in small cell enhancements in LTE advanced}', \emph{IEEE
  Commun Mag},  2013, \textbf{51}, (2), pp.~98--105

\bibitem{al2017full}
Al.Kadri, M.O., Deng, Y., Aijaz, A., Nallanathan, A.: `{Full-Duplex Small Cells
  for Next Generation Heterogeneous Cellular Networks: A Case Study of Outage
  and Rate Coverage Analysis}', \emph{{IEEE Access}},  2017, \textbf{5},
  pp.~8025--8038

\bibitem{galkin2017stochastic}
Galkin, B., Kibi{\l}da, J., DaSilva, L.A.: `A stochastic geometry model of
  backhaul and user coverage in urban uav networks', \emph{arXiv preprint
  arXiv:171003701},  2017,

\bibitem{turgut2018downlink}
Turgut, E., Gursoy, M.C.: `{Downlink Analysis in Unmanned Aerial Vehicle (UAV)
  Assisted Cellular Networks with Clustered Users}', \emph{IEEE Access},  2018,
  \textbf{6}, pp.~36313--36324

\bibitem{azari2017coexistence}
Azari, M.M., Rosas, F., Chiumento, A., Pollin, S.
\newblock `Coexistence of terrestrial and aerial users in cellular networks'.
\newblock In: Proc. IEEE Global Commun. Conf. (Singapore,  2017. pp.~ 1--6

\bibitem{singh2019distributed}
Singh, S., Kumbhar, A., G{\"u}ven{\c{c}}, I., Sichitiu, M.L.
\newblock `{Distributed Approaches for Inter-cell Interference Coordination in
  UAV-based LTE-Advanced HetNets}'.
\newblock In: Proc. IEEE 88th Vehic. Technol. Conf. (VTC-Fall). (Chicago, IL,
  2019. pp.~ 1--6

\bibitem{naranguav}
Narang, M., Xiang, S., Liu, W., Gutierrez, J., Chiaraviglio, L., Sathiaseelan,
  A., et~al.
\newblock `{UAV}-assisted edge infrastructure for challenged networks'.
\newblock In: Proc. IEEE Conf. Comp. Commun. Workshops (INFOCOM Workshop).
  (Atlanta, GA,  2017. pp.~ 60--65

\bibitem{al2014optimal}
Al.Hourani, A., Kandeepan, S., Lardner, S.: `{Optimal LAP altitude for maximum
  coverage}', \emph{IEEE Wireless Commun Lett},  2014, \textbf{3}, (6),
  pp.~569--572

\bibitem{bor2016efficient}
Bor.Yaliniz, R.I., El.Keyi, A., Yanikomeroglu, H.
\newblock `Efficient {3-D} placement of an aerial base station in next
  generation cellular networks'.
\newblock In: Proc. IEEE Intl. Conf. Commun. (ICC). (Kuala Lumpur, Malaysia,
  2016. pp.~ 1--5

\bibitem{mozaffari2016optimal}
Mozaffari, M., Saad, W., Bennis, M., Debbah, M.
\newblock `Optimal transport theory for power-efficient deployment of unmanned
  aerial vehicles'.
\newblock In: Proc. IEEE Intl. Conf. Commun. (ICC). (Kuala Lumpur, Malaysia,
  2016. pp.~ 1--6

\bibitem{sharma2016uav}
Sharma, V., Bennis, M., Kumar, R.: `{UAV-assisted} heterogeneous networks for
  capacity enhancement', \emph{IEEE Commun Lett},  2016, \textbf{20}, (6),
  pp.~1207--1210

\bibitem{christy2017optimum}
Christy, E., Astuti, R.P., Syihabuddin, B., Narottama, B., Rhesa, O.,
  Rachmawati, F.
\newblock `{Optimum UAV flying path for Device-to-Device communications in
  disaster area}'.
\newblock In: Proc. IEEE Int. Conf. Sig Sys. (ICSigSys). (Bali, Indonesia,
  2017. pp.~ 318--322

\bibitem{rupasinghe2019non}
Rupasinghe, N., Yap{\i}c{\i}, Y., G{\"u}ven{\c{c}}, I., Kakishima, Y.:
  `Non-orthogonal multiple access for mmwave drone networks with limited
  feedback', \emph{IEEE Trans Commun},  2019, \textbf{67}, (1), pp.~762--777

\bibitem{sun2018location}
Sun, Y., Wang, T., Wang, S.
\newblock `{Location Optimization for Unmanned Aerial Vehicles Assisted Mobile
  Networks}'.
\newblock In: Proc. IEEE Intl. Conf. Commun. (ICC). (Kansas City, MO,  2018.
  pp.~ 1--6

\bibitem{hanna2019distributed}
Hanna, S., Yan, H., Cabric, D.
\newblock `{Distributed UAV Placement Optimization for Cooperative
  Line-of-sight MIMO Communications}'.
\newblock In: Proc. IEEE Intl. Conf. Acoustics, Speech, Signal Process.
  (ICASSP). (Brighton, UK,  2019. pp.~ 4619--4623

\bibitem{challita2019interference}
Challita, U., Saad, W., Bettstetter, C.: `{Interference Management for
  Cellular-Connected UAVs: A Deep Reinforcement Learning Approach}', \emph{IEEE
  Trans Wireless Commun},  2019, \textbf{18}, (4), pp.~2125--2140

\bibitem{mozaffari2015drone}
Mozaffari, M., Saad, W., Bennis, M., Debbah, M.
\newblock `Drone small cells in the clouds: Design, deployment and performance
  analysis'.
\newblock In: Proc. IEEE Global Commun. Conf. (GLOBECOM). (San Diego, CA,
  2015. pp.~ 1--6

\bibitem{saquib2013fractional}
Saquib, N., Hossain, E., Kim, D.I.: `Fractional frequency reuse for
  interference management in {LTE-advanced} hetnets', \emph{IEEE Wireless
  Commun},  2013, \textbf{20}, (2), pp.~113--122

\bibitem{kaleem2016public}
Kaleem, Z., Chang, K.: `{Public Safety Priority-Based User Association for Load
  Balancing and Interference Reduction in PS-LTE Systems}', \emph{IEEE Access},
   2016, \textbf{4}, pp.~9775--9785

\bibitem{deb2014algorithms}
Deb, S., Monogioudis, P., Miernik, J., Seymour, J.P.: `Algorithms for enhanced
  inter-cell interference coordination {(eICIC)} in {LTE} hetnets',
  \emph{IEEE/ACM Trans Nwk},  2014, \textbf{22}, (1), pp.~137--150

\bibitem{R10}
Merwaday, A., Mukherjee, S., G{\"u}ven{\c{c}}, I.: `{Capacity analysis of
  LTE-Advanced HetNets with reduced power subframes and range expansion}',
  \emph{EURASIP J Wireless Commun Netw},  2014, \textbf{2014}, (1), pp.~1--19

\bibitem{mukherjee2011effects}
Mukherjee, S., G{\"u}ven{\c{c}}, I.
\newblock `Effects of range expansion and interference coordination on capacity
  and fairness in heterogeneous networks'.
\newblock In: Proc. IEEE Asilomar Conf. Signals, Systems and Computers.
  (Pacific Grove, CA, USA,  2011. pp.~ 1855--1859

\bibitem{kumbhar2018interference}
Kumbhar, A., Binol, H., Guvenc, I., Akkaya, K.
\newblock `{Interference Coordination for Aerial and Terrestrial Nodes in
  Three-Tier LTE-Advanced HetNet}'.
\newblock In: Proc. IEEE Radio Wireless Symp. (RWS). (Orlando, FL,  2019. pp.~
  1--4

\bibitem{zhang2018machine}
Zhang, Q., Mozaffari, M., Saad, W., Bennis, M., Debbah, M.
\newblock `{Machine learning for predictive on-demand deployment of UAVs for
  wireless communications}'.
\newblock In: Proc. IEEE Global Commun. Conf. (GLOBECOM). (Abu Dhabi, United
  Arab Emirates,  2018. pp.~ 1--6

\bibitem{zhang2018predictive}
Zhang, Q., Saad, W., Bennis, M., Lu, X., Debbah, M., Zuo, W.: `{Predictive
  Deployment of UAV Base Stations in Wireless Networks: Machine Learning Meets
  Contract Theory}', \emph{arXiv preprint arXiv:181101149},  2018,

\bibitem{sharafeddine2019demand}
Sharafeddine, S., Islambouli, R.: `On-demand deployment of multiple aerial base
  stations for traffic offloading and network recovery', \emph{Elsevier
  Computer Networks},  2019, \textbf{156}, pp.~52--61

\bibitem{li2018placement}
Li, P., Xu, J.: `{Placement Optimization for UAV-Enabled Wireless Networks with
  Multi-Hop Backhauls}', \emph{Springer J Commun Information Netw},  2018,
  \textbf{3}, (4), pp.~64--73

\bibitem{fouda2019interference}
Fouda, A., Ibrahim, A.S., G{\"u}ven{\c{c}}, {\'I}., Ghosh, M.: `{Interference
  management in uav-assisted integrated access and backhaul cellular
  networks}', \emph{IEEE Access},  2019, \textbf{7}, pp.~104553--104566

\bibitem{drive2}
{The Drive}.
\newblock `{Drone Saves Man's Life From Kilauea Volcano Disaster in Hawaii}'.
\newblock (,  2018.
\newblock Available from:
  \url{http://www.thedrive.com/tech/21276/drone-saves-mans-life-from-kilauea-volcano-disaster-in-hawaii}

\bibitem{menouar2017uav}
Menouar, H., Guvenc, I., Akkaya, K., Uluagac, A.S., Kadri, A., Tuncer, A.:
  `{UAV-enabled intelligent transportation systems for the smart city:
  Applications and challenges}', \emph{IEEE Commun Mag},  2017, \textbf{55},
  (3), pp.~22--28

\bibitem{koohifar2017receding}
Koohifar, F., Kumbhar, A., Guvenc, I.: `Receding horizon multi-uav cooperative
  tracking of moving rf source', \emph{IEEE Commun Let},  2017, \textbf{21},
  (6), pp.~1433--1436

\bibitem{saputro2018supporting}
Saputro, N., Akkaya, K., Uluagac, S.
\newblock `{Supporting Seamless Connectivity in Drone-assisted Intelligent
  Transportation Systems}'.
\newblock In: Proc. IEEE Local Computer Networks Workshops (LCN Workshops).
  (Chicago, IL,  2018. pp.~ 110--116

\bibitem{niu2018uav}
Niu, H., Gonzalez.Prelcic, N., Heath, R.W.
\newblock `A uav-based traffic monitoring system-invited paper'.
\newblock In: Proc. IEEE Vehic. Tech. Conf. (VTC Spring). (Porto, Portugal,
  2018. pp.~ 1--5

\bibitem{van2016lte}
Van~der Bergh, B., Chiumento, A., Pollin, S.: `{LTE in the sky: Trading off
  propagation benefits with interference costs for aerial nodes}', \emph{IEEE
  Commun Mag},  2016, \textbf{54}, (5), pp.~44--50

\bibitem{amorim2018measured}
Amorim, R., Nguyen, H., Wigard, J., Kov{\'a}cs, I.Z., S{\o}rensen, T.B., Biro,
  D.Z., et~al.: `{Measured uplink interference caused by aerial vehicles in LTE
  cellular networks}', \emph{IEEE Wireless Commun Let},  2018, \textbf{7}, (6),
  pp.~958--961

\bibitem{liu2018performance}
Liu, C., Ding, M., Ma, C., Li, Q., Lin, Z., Liang, Y.C.
\newblock `{Performance analysis for practical unmanned aerial vehicle networks
  with LoS/NLoS transmissions}'.
\newblock In: Proc. IEEE Intl. Conf. Commun. Workshops (ICC Workshops). (Kansas
  City, MO,  2018. pp.~ 1--6

\bibitem{3GPP.TR.36.873dup}
{3GPP}.
\newblock `{Study on 3D channel model for LTE (Release 12)}'.
\newblock ({3rd Generation Partnership Project (3GPP)},  2017. 36.873.
\newblock version 12.7.0

\bibitem{kammoun2014preliminary}
Kammoun, A., Khanfir, H., Altman, Z., Debbah, M., Kamoun, M.: `{Preliminary
  results on 3D channel modeling: From theory to standard}', \emph{IEEE J Sel
  Areas Commun (JSAC)},  2014, \textbf{32}, (6), pp.~1219--1229

\bibitem{xiroOnline}
{Xiro Online}.
\newblock `Okumura-{H}ata'.
\newblock (,  2017.
\newblock Available from:
  \url{https://www.xirio-online.com/help/en/okumurahata.html}

\bibitem{3GPP.TR.36.777}
3GPP.
\newblock `{Study on Enhanced LTE Support for Aerial Vehicles (Release 15)}'.
\newblock ({3rd Generation Partnership Project (3GPP)},  2017. 36.777.
\newblock version 15.0.0

\bibitem{ITU-R.P.1410-2}
ITU.
\newblock `{Propagation data and prediction methods required for the design of
  terrestrial broadband millimetric radio access systems}'.
\newblock ({International Telecommunication Union (ITU)},  2003. P.1410-2

\bibitem{khawaja2018survey}
Khawaja, W., Guvenc, I., Matolak, D., Fiebig, U.C., Schneckenberger, N.: `A
  survey of air-to-ground propagation channel modeling for unmanned aerial
  vehicles', \emph{IEEE Commun Survey Tuts},  2019, \textbf{21}, pp.~2361--2391

\bibitem{binol2018hybrid}
Binol, H., Guvenc, I., Bulut, E., Akkaya, K.: `Hybrid evolutionary search
  method for complex function optimisation problems', \emph{IET Electronics
  Letters},  2018, \textbf{54}, (24), pp.~1377--1379

\bibitem{binol2018time}
Binol, H., Bulut, E., Akkaya, K., Guvenc, I.
\newblock `Time optimal multi-uav path planning for gathering its data from
  roadside units'.
\newblock In: Proc. IEEE 88th Vehicular Technology Conference (VTC-Fall).
  (Chicago, IL,  2018. pp.~ 1--5

\bibitem{moore2014first}
Moore, L.K.: `The first responder network (FirstNet) and next-generation
  communications for public safety: Issues for congress'.
\newblock (Congressional Research Service,  2014)

\end{thebibliography}

\vfill\pagebreak

\end{document}